\documentclass[lettersize,journal]{IEEEtran}
\usepackage{amsmath,amsfonts}
\usepackage{algorithmic}
\usepackage{algorithm}
\usepackage{amssymb}
\usepackage{array}
\usepackage{graphicx}
\usepackage{textcomp}
\usepackage{stfloats}
\usepackage{makecell}
\usepackage{url}
\usepackage{upgreek}
\usepackage{verbatim}
\usepackage[caption=false,font=footnotesize]{subfig}

\usepackage{colortbl}
\usepackage{varwidth}
\usepackage{microtype}
\usepackage{booktabs}
\usepackage{multirow}
\usepackage{capt-of}
\def\BibTeX{{\rm B\kern-.05em{\sc i\kern-.025em b}\kern-.08em
		T\kern-.1667em\lower.7ex\hbox{E}\kern-.125emX}}
\usepackage{balance}

\usepackage{cite}
\begin{document}

\title{UAV SAR Imaging via Diffusion Posterior Sampling\\
	under Non-Ideal Motion in ISAC Systems}

\author{Keying Zhu, \textsl{Graduate Student Member, IEEE}, Le Liang, \textsl{Member, IEEE}, Jie Yang, \textsl{Member, IEEE},\\ Fan Liu, \textsl{Member, IEEE}, and Shi Jin, \textsl{Fellow, IEEE}
\thanks{Keying Zhu and Fan Liu are with the National Mobile Communications Research Laboratory, Southeast University, Nanjing 210096, China (e-mail: keyingz@seu.edu.cn; fan.liu@seu.edu.cn).}
\thanks{Le Liang is with the National Mobile Communications Research Laboratory, Southeast University, Nanjing 210096, China, and also with Purple Mountain Laboratories, Nanjing 211111, China (e-mail: lliang@seu.edu.cn).}
\thanks{Jie Yang and Shi Jin are with the National Mobile Communications Research Laboratory and the Frontiers Science Center for Mobile Information Communication and Security, Southeast University, Nanjing 210096, China (e-mail: yangjie@seu.edu.cn; jinshi@seu.edu.cn).}

}

\markboth{Journal of \LaTeX\ Class Files,~Vol.~14, No.~8, August~2021}
{Shell \MakeLowercase{\textit{et al.}}: A Sample Article Using IEEEtran.cls for IEEE Journals}


\maketitle
\raggedbottom

\begin{abstract}
	Unmanned aerial vehicle-enabled synthetic aperture radar (SAR) offers significant potential for providing communication services to ground users while sensing the surrounding environment in low-altitude integrated sensing and communication (ISAC) systems. However, its practical deployment is challenged by known non-ULM trajectories and unknown position errors, which violate conventional SAR assumptions and cause severe image defocusing.
	We address these non-ideal impacts by formulating the ISAC SAR imaging as an exact linear inverse problem to accurately model arbitrary trajectories with multiple-input multiple-output orthogonal frequency division multiplexing waveforms. Based on this formulation, we develop a diffusion posterior sampling (DPS)-based super-resolution SAR framework, where a pre-trained diffusion model provides a data-driven image prior and the exact physical model enforces measurement consistency during posterior inference. The inference process is further accelerated for practical implementation. 
	For unknown position errors, we develop a joint iterative DPS and motion compensation (MoCo) optimization algorithm with discrete cosine transform reparameterization and coarse-to-fine spectral scheduling. Simulations demonstrate that our framework achieves high-accuracy reconstruction across diverse trajectories and remarkable super-resolution, maintaining greater than 21.3 dB peak signal-to-noise ratio (PSNR) even under half physical resolution truncation. Under unknown motion errors, the joint DPS and MoCo algorithm achieves 23.65 dB PSNR with a residual position-error estimation error of 0.19 cm.
\end{abstract}

\begin{IEEEkeywords}
	Integrated sensing and communication, diffusion posterior sampling, synthetic aperture radar, non-ideal trajectory
\end{IEEEkeywords}

\vspace{-2mm}
\section{Introduction}
\IEEEPARstart{I}{NTEGRATED} sensing and communications (ISAC) has emerged as a critical technology for the sixth generation (6G) wireless systems, enabling joint high-quality wireless connectivity and high-accuracy environment perception via shared spectral, hardware, and algorithmic resources \cite{LF_survey,Azhangsurvey}. Concurrently, the rapid development of the low-altitude economy extends ISAC into the aerial domain. Unmanned aerial vehicles (UAVs) serve as aerial dual-functional base stations, simultaneously delivering high-throughput communication services and sensing the complex surrounding environment, offering critical technical support for key missions like urban surveillance and disaster relief \cite{LF_UAVsurvey,huang_imaging}.
These applications impose higher communication and sensing requirements on low-altitude ISAC systems. Although the multiple-input multiple-output (MIMO) and orthogonal frequency division multiplexing (OFDM) modules carried by the UAV enable high-throughput communication, the limited physical antenna aperture constrains the achievable spatial resolution, making it insufficient for high-quality scene imaging \cite{zhu2025super}. However, the inherent mobility of UAVs offers a compelling solution. By leveraging synthetic aperture radar (SAR) principles \cite{SAR_survey}, backscattered ISAC echoes collected along the flight trajectory can be coherently integrated to synthesize a large virtual antenna array, thereby overcoming the hardware constraints to achieve high-resolution imaging.

While UAV SAR offers immense potential, its practical implementation in low‑altitude ISAC systems introduces unprecedented signal processing challenges. Unlike traditional SAR in the radar domain, which rely on a specially designed ideal uniform linear motion (ULM) trajectory, ISAC UAVs must execute highly flexible three-dimensional (3D) trajectories to jointly optimize communication link quality and sensing coverage. 
Furthermore, the imaging process relies on OFDM communication echoes rather than dedicated radar pulses. Consequently, under non-ideal flexible trajectories, the target slant range becomes deeply coupled with pixel spatial positions, slow-time indices, array geometry, and subcarrier frequencies. 
This intricate coupling induces severe 3D spatial variance and nonlinear Doppler modulations during the ISAC SAR imaging process, thereby rendering classical SAR algorithms like chirp scaling algorithms \cite{chirp_scaling} and range-Doppler algorithm (RDA) \cite{RDA} ineffective, as their fundamental reliance on ULM assumptions and decoupled spatial approximations means that direct application to flexible UAV trajectories inevitably leads to catastrophic range defocusing and azimuth smearing.

Time-domain back projection (BP) algorithms \cite{BP} partially alleviate these challenges by computing exact geometric distances, while it strictly requires well-conditioned data.  Under communication-induced undersampling in practical ISAC scenarios, BP inevitably suffers from severe sidelobes and spatial artifacts. To circumvent this limitation, compressed sensing formulates SAR as a linear inverse problem with sparse prior \cite{CS_SAR1,CS_SAR2,CS_SAR3}. Unfortunately, these handcrafted priors fail to adequately capture the complex textures and spatial structures of real SAR scenes. Critically, pursuing super-resolution with employing finer imaging grids drastically increases the column coherence of the forward measurement matrix, which violates the restricted isometry property (RIP) required by compressed sensing \cite{rip}, leading to algorithm divergence and severely limiting super-resolution capability in ISAC systems.

Driven by the integration of artificial intelligence into ISAC, deep learning has been widely adopted in SAR imaging to extract highly expressive representations \cite{DL_SAR_survey}. Existing deep learning based methods generally adopt a supervised learning paradigm to directly map echoes to SAR images \cite{DL_SAR2}. However, such mappings are coupled with the sensing geometry encountered during training and therefore generalize poorly to flexible UAV trajectories, MIMO configurations, and OFDM sampling strategies \cite{DL_SAR1}. 
To achieve generalizable SAR imaging without prohibitive scenario-specific retraining, decoupling the physical observation model from image priors is essential. Although plug-and-play (PnP) frameworks offer such decoupling \cite{pnp}, conventional denoisers struggle to capture the complex heavy-tailed distributions of real SAR scenes, ultimately limiting reconstruction fidelity.
Recently, generative diffusion models, including the denoising diffusion probabilistic model (DDPM) \cite{DDPM} and score-based generative models \cite{scorebased}, have emerged as a transformative paradigm, demonstrating an exceptional capacity to capture highly intricate data distributions and synthesize diverse, high-quality samples through straightforward training procedures \cite{rombach2022high}. 
Recently, extensive research has explored the potential of diffusion models in solving linear inverse problems. Diffusion posterior sampling (DPS) \cite{DPS} acts a pre-trained diffusion model as an offline, universal spatial prior, while the physical observations serve as the online likelihood guidance, thereby completely bypassing the retraining dilemma \cite{DPS_MRI}. However, while this decoupled generative inference paradigm has been widely investigated in fields such as high-dimensional channel estimation \cite{zhou_DM,CE_score}, its exploration and adaptation for highly dynamic, ISAC SAR super-resolution imaging system remain largely uncharted.

Despite the theoretical elegance of generative diffusion models, successfully deploying it in realistic low-altitude ISAC systems presents an additional challenge. In practice, light-weight UAVs inevitably suffer from unknown motion errors due to wind or rotor vibrations during flight \cite{moco_survey,moco1,moco2}. These unknown flight perturbations introduce severe phase errors into the echoes, yielding an inaccurate measurement model and leading to drastic image defocusing. Traditional motion errors compensation (MoCo) algorithms, such as phase gradient autofocus \cite{PGA}, are primarily designed for conventional SAR imaging under the ULM assumption. They are fundamentally incompatible in low-altitude ISAC systems and may even introduce erroneous autofocusing phases. Therefore, a joint framework capable of simultaneously compensating for unknown trajectory errors and reconstructing the super-resolution image is urgently desired.

Based on the above discussion, this paper proposes a novel UAV SAR imaging algorithmic framework tailored for low‑altitude ISAC systems. The framework integrates DPS with iterative MoCo to achieve robust super‑resolution imaging under arbitrary known trajectories, MIMO‑OFDM waveforms, and unknown flight perturbations. The main contributions of this paper are summarized as follows:
\begin{itemize}
	\item \textbf{Generalized ISAC SAR Model and Non-Ideal Motion Analysis}: We establish a generalized UAV SAR echo model tailored for ISAC systems, exactly integrating arbitrary 3D trajectories, MIMO array geometries, and OFDM waveforms. Based on this, we analyze the impacts of two critical non-ideal motion factors on conventional SAR algorithms: the known non-ULM reference trajectories and the unknown position errors. This analysis lays a theoretical foundation for our proposed robust framework.

	\item \textbf{DPS-Based Super-Resolution SAR Imaging under Known Trajectories}: For known reference trajectories, we formulate ISAC SAR imaging as an exact linear inverse problem, entirely bypassing conventional decoupling approximations. A pre-trained diffusion model is introduced as a data-driven prior, while the echoes provide online likelihood guidance through DPS, enabling high-fidelity and super-resolution imaging across arbitrary trajectories without retraining. 
	Moreover, we incorporate the denoising diffusion implicit model (DDIM) \cite{DDIM} acceleration mechanism to reduce the inference cost.

	\item \textbf{Joint DPS and MoCo under Unknown Position Errors}: 
	We develop a joint iterative DPS and MoCo framework that alternately refines the SAR image and the position errors throughout the reverse diffusion process to compensate for unknown UAV motion perturbations. By incorporating discrete cosine transform (DCT) reparameterization and a coarse-to-fine spectral scheduling strategy, our method stabilizes the joint estimation, ensuring robust, high-fidelity imaging under severely non-ideal motion.
\end{itemize}

\textsl{Notations}: Boldface letters $\mathbf{a}$ and $\mathbf{A}$ denote a vector and a matrix, respectively. 
$\mathbf{A}^{\text{T}}$, $\mathbf{A}^{\text{H}}$, and $\mathbf{A}^{-1}$ denote the transpose, conjugate transpose, and inverse, respectively.
$\|\mathbf{a}\|_{2} \triangleq (\mathbf{a}^{\text{H}}\mathbf{a})^{1/2}$ denotes Euclidean norm of $\mathbf{a}$, and $\|\mathbf{a}\|_{2}^{2}$ denotes the corresponding squared Euclidean norm.
$\odot$ denote the Hadamard product.  
$\nabla_{x}f(x)$ denotes the gradient of $f(\cdot)$ with respect to $x$. 
$\mathbb{E}[\cdot]$ represents the statistical expectation or the expectation operator. $\mathbf{I}$ denotes the identity matrix. $\mathbb{R}$ and $\mathbb{C}$ are the sets of real and complex numbers, respectively.

\begin{figure}
	\centering
	\includegraphics[width=0.48\textwidth]{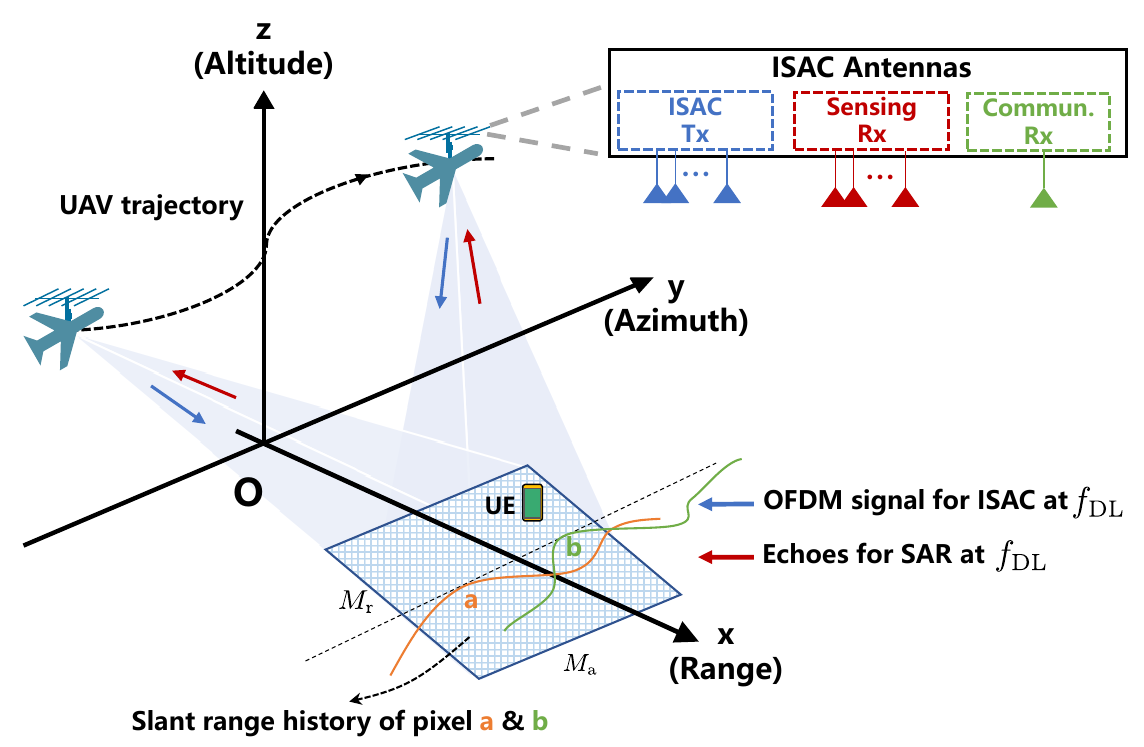}
	\caption{Framework of UAV SAR imaging in ISAC system.}
	\label{fig:system_model}
	\vspace{-3mm}
\end{figure}

\section{System Model And Problem Formulation}
In a low-altitude ISAC system, we consider a UAV deployed as a dual-functional aerial base station integrated with communication and sensing capabilities, as illustrated in Fig. \ref{fig:system_model}. 
The UAV concurrently provides downlink OFDM communication services to ground users and exploits its mobility to perform high-resolution SAR imaging using the backscattered communication echoes. 
However, unlike traditional stable platforms, the highly flexible trajectories and inherent flight instability of UAVs, deeply coupled with the ISAC MIMO-OFDM waveforms, induce complex phase perturbations that severely degrade imaging quality.
Therefore, this section establishes a generalized echo signal model and mathematically analyzes the impacts of non-ideal UAV motion, thereby exposing the limitations of conventional SAR algorithms and laying the physical groundwork for our proposed imaging framework.

\vspace{-1mm}
\subsection{Geometric Model}
We assume the UAV adopts a frequency division duplexing architecture with a hybrid antenna configuration aligned along the longitudinal axis of the UAV to support the ISAC dual-functionalities. As shown in Fig. \ref{fig:system_model}, an ISAC transmit (Tx) uniform linear array (ULA) comprising $N_{\text{t}}$ elements is utilized to broadcast downlink OFDM signals for both communication and sensing in the $f_{\mathrm{DL}}$ band. 
Alongside the ISAC Tx array, a dedicated sensing receive (Sensing Rx) ULA consisting of $N_{\text{r}}$ elements, captures backscattered echoes within the same band. For both ULAs, the antenna spacing $d$ is set to $\lambda/2$, where $\lambda$ is the wavelength. To ensure that the sensing process remains immune to uplink interference, uplink communications are handled by a separate single-antenna receiver (Commun. Rx) operating in a non-overlapping band $f_{\mathrm{UL}}$. Furthermore, to suppress self-interference, the ISAC Tx and the Sensing Rx arrays are separated by a distance $D_{\mathrm{off}}$ of approximately 5$\lambda$ to 10$\lambda$.
By leveraging multi-antenna digital beamforming, the UAV achieves precise steering for robust downlink communication services, which is naturally aligned with the spotlight SAR mode, as it maintains energy focus on the target area throughout the synthetic aperture duration. Such synergy facilitates simultaneous stable communication and high-resolution sensing via efficient energy aggregation.

We establish a generalized 3D instantaneous trajectory model to characterize the high mobility flight of UAVs. Considering the non-uniform velocity fluctuations and trajectory changes of UAVs, the instantaneous velocity and position of the UAV's platform center are defined as $\mathbf{v}(t)=[v_{\mathrm{x}}(t),v_{\mathrm{y}}(t),v_{\mathrm{z}}(t)]^{\mathrm{T}}$ and $\mathbf{p}(t) = [x(t), y(t), z(t)]^{\mathrm{T}}$, respectively. 
Given the physical layout of the ISAC arrays, the instantaneous positions of the $p$-th ($p=1,\dots,N_{\mathrm{t}}$) Tx element and the $q$-th ($q=1,\dots,N_{\mathrm{r}}$) Sensing Rx element are represented by coupling translational motion with rotational attitude
\begin{equation}
	\mathbf{p}_{*}(t) = \mathbf{p}(t) + \mathbf{R}(t)\mathbf{d}_{*},* \in \{p,q\},
	\label{II-A-eq:position}
\end{equation}
where $\mathbf{R}(t) \in \mathbb{R}^{3\times 3}$ is the rotation matrix \cite{rotation} that transforms the antenna coordinates from the body-fixed frame to the global coordinate system.
The body-fixed offset vectors $\mathbf{d}_{p}$ and $\mathbf{d}_{q}$ are determined by the physical ULAs geometry as
	\begin{equation}
		\begin{aligned}
			& \mathbf{d}_{p}=
			\Big[0, \ \big(p-\dfrac{N_{\mathrm{t}}+1}{2}\big)d
			-\dfrac{D_{\mathrm{off}}}{2},\ 0 \Big]^{\mathrm{T}}, \\
			&\mathbf{d}_{q}=
			\Big[0, \ \big(q-\dfrac{N_{\mathrm{r}}+1}{2}\big)d
			+\dfrac{D_{\mathrm{off}}}{2},\ 0 \Big]^{\mathrm{T}}.
		\end{aligned}
		\label{eq:slant_range}
	\end{equation}
For analytical tractability, the target imaging region is uniformly discretized into $M= M_{\mathrm{r}} \times M_{\mathrm{a}}$ pixels. The coordinates of the $m$-th ($m=1,\dots,M$) pixel are denoted by $\mathbf{t}_m=[x_m,y_m,0]^{\mathrm{T}}$. 
Consequently, the Tx-target-Rx total propagation distance for the $m$-th pixel associated with the $(p,q)$ antenna pair is  
	\begin{equation}
		R_m^{p,q}(t)
		=
		\left\|\mathbf{p}_p(t)-\mathbf{t}_m\right\|_2
		+
		\left\|\mathbf{p}_q(t)-\mathbf{t}_m\right\|_2,
		\label{eq:slant_range}
	\end{equation}
where the two terms represent the one-way slant ranges from the $p$-th Tx element to the $m$-th pixel and from the $m$-th pixel to the $q$-th sensing Rx element, respectively.

\vspace{-1mm}
\subsection{Signal Model}
As shown in Fig. \ref{fig:system_model}, the UAV successively transmits downlink OFDM signals containing $K$ subcarriers and $U$ symbols at a carrier frequency $f_{\mathrm{c}}$ (within band $f_{\mathrm{DL}}$) with a subcarrier spacing of $\Delta f$ throughout its trajectory. The total OFDM symbol period is $T_{\mathrm{s}}=T_{\mathrm{cp}}+T_{\mathrm{d}}$, where $T_{\mathrm{d}}$ is the data symbol duration equaling $1/\Delta f$, and $T_{\mathrm{cp}}$ is the length of the cyclic prefix (CP). To effectively suppress inter-symbol interference and accommodate the imaging swath, the CP duration is chosen to satisfy $T_{\mathrm{cp}} \ge (R_{\mathrm{far}}-R_{\mathrm{near}})/c$, where $R_{\mathrm{far}}$ and $R_{\mathrm{near}}$ denote the maximum and minimum Tx-target-Rx total propagation distances of the target region during flight, respectively \cite{OFDM_CP}. $c$ is the speed of light. The baseband OFDM signal emitted from the $p$-th Tx element is
\begin{equation}
	s^p(t) \! =\! \frac{1}{\sqrt{K}}\! \sum_{u=0}^{U-1}\! \sum_{k=0}^{K-1} d_{u,k}^p e^{j 2\pi k \Delta f (t - u T_{\mathrm{s}})} \mathrm{rect}\Big(\frac{t- u T_{\mathrm{s}}}{T_{\mathrm{s}}}\Big),
	\label{II-B-eq:transmit signal}
\end{equation}
where $d_{u,k}^p$ denotes the complex frequency-domain coefficient transmitted by the $p$-th Tx element over the $k$-th subcarrier of the $u$-th OFDM symbol after data modulation and spatial precoding. Crucially, $\{d_{u,k}^p\}_{p=1}^{N_{\text{t}}}$ jointly form the MIMO transmit vector, whose spatial precoding serves communication users while providing the required sensing illumination throughout the flight.
The function $\mathrm{rect}(t/T)$ depicts a rectangular pulse equaling one for $t \in [0,T]$ and zero otherwise. 

To characterize the SAR echoes, let $\eta$ and $t_f$ denote the continuous slow-time variable along the synthetic aperture in the azimuth dimension and the continuous fast-time variable in the range dimension, respectively. The SAR echoes are sampled at $N_{\mathrm{sa}}$ discrete slow-time instants $\eta_{\ell}$, where $\ell=0,\dots,N_{\mathrm{sa}}-1$ is the slow-time sample index.
Enabled by the spatial separation between the ISAC Tx and Sensing Rx, the UAV can simultaneously transmit the ISAC waveform and receive the corresponding SAR echoes. 
Since the OFDM symbol rate $1/T_{\mathrm{s}}$ significantly exceeds the slow-time sampling rate required for SAR imaging, processing every OFDM symbol is highly redundant. 
Consequently, the receiver retains one OFDM echo every $Q_{\rm sym}$ symbols over the synthetic aperture duration $T_{\mathrm{sa}}$, yielding $N_{\mathrm{sa}}$ slow-time samples.
The resulting equivalent pulse repetition frequencies (PRF) is 
\begin{equation}
  f_{\text{PRF}}^{\text{eq}}=\frac{1}{\eta_{\ell+1}-\eta_{\ell}}=\frac{1}{Q_{\mathrm{sym}}T_{\mathrm{s}}}.
  \label{II-B-eq:PRF}
\end{equation}
Let $u_{\ell}$ denote the OFDM symbol associated with the $\ell$-th slow-time sample. The corresponding fast-time baseband waveform at the $p$-th ISAC Tx element follows from (\ref{II-B-eq:transmit signal}) as
\begin{equation}
	s_{\ell}^p(t_f) =\frac{1}{\sqrt{K}}\sum_{k=0}^{K-1} d_{u_{\ell},k}^p e^{j 2\pi k \Delta f t_f}, \quad 0 \leq t_f \leq T_{\text{s}}, 
\end{equation}
Accordingly, the complex baseband echo received by the $q$-th Sensing Rx element at $\eta_{\ell}$ is
\begin{equation}
	\begin{aligned}
		y^{q}(t_f, \eta_{\ell}) = \sum_{p=1}^{N_{\mathrm{t}}}  \sum_{m=1}^{M} &A_m^{p,q}(\eta_{\ell}) \chi_m s^p_{\ell}\big(t_f - \tau_m^{p,q}  (\eta_{\ell}) \big) \\[-1ex]
		&\times e^{-j 2\pi f_\mathrm{c} \tau_m^{p,q}(\eta_{\ell})} + w^{q}(t_f,\eta_{\ell}),
	\end{aligned}
	\label{eq:time_domain_echo}
\end{equation}
where $\chi_m$ is the unknown complex reflectivity coefficient of the $m$-th pixel, and $A_m^{p,q}(\eta_{\ell})$ denotes the known amplitude response associated with the $m$-th pixel and the $(p,q)$ antenna pair at slow time $\eta_{\ell}$, accounting for the propagation loss, antenna pattern, illumination gain, and other system-dependent factors.
$\tau_m^{p,q}(\eta_{\ell})=R_m^{p,q}(\eta_{\ell})/c $ denotes the total propagation delay determined by the propagation geometry in (\ref{eq:slant_range}), and $w^{q}(t_f,\eta_{\ell})$ is the additive noise.
At the receiver, the echoes in (\ref{eq:time_domain_echo}) are first processed by CP removal, followed by a $K$-point fast Fourier transform along the fast-time dimension.
The resulting frequency-domain observation is given by
\begin{equation}
	\begin{aligned}
		y^{q}[k, \ell] \!=\!\! \sum_{p=1}^{N_{\mathrm{t}}}\! \sum_{m=1}^{M}\! A_{m}^{p, q}[k, \ell] d_{u_{\ell}, k}^{p} \chi_{m} &e^{-j 2 \pi (f_{\mathrm{c}} + k \Delta f) {\tau}_{m}^{p, q}(\eta_{\ell})} \\[-2ex] 
		&\quad + w^{q}[k, \ell],
	\end{aligned}
	\label{II-B-eq:RF_echo_model}
\end{equation}
where $A_m^{p,q}[k,\ell]$ is the frequency-domain amplitude response, and $w^{q}[k, \ell]$ is the frequency-domain equivalent additive white Gaussian noise. The ISAC SAR imaging task is to recover the complex reflectivity map $\boldsymbol{\chi}=[\chi_1,\dots,\chi_{M}]^{\text{T}}$ from (\ref{II-B-eq:RF_echo_model}), which characterizes the scattering properties of the target area.

\vspace{-1mm}
\subsection{Impacts of Non-Ideal Motion}
Conventional SAR imaging algorithms like RDA are typically developed under the ULM assumption, which enables approximate decoupling of the range and azimuth processing.
However, in low-altitude ISAC systems, the UAV follows a flexible 3D trajectory to accommodate communication coverage and sensing perspectives, while being further affected by unknown position perturbations. 
To characterize these two forms of non-ideal motion, let $\mathbf p^{\mathrm{ref}}(\eta)$ and $\mathbf p^{\mathrm{real}}(\eta)$ denote the known reference and actual trajectories of the UAV platform center along the SAR slow-time dimension, respectively. Then, 
\vspace{-1mm}
\begin{equation}
	\mathbf{p}^{\mathrm{real}}(\eta) = \mathbf{p}^{\mathrm{ref}}(\eta) + \Delta\mathbf{p}(\eta),
	\label{II-C-eq:center_real_ref_trajectory}
\end{equation}
where $\mathbf{p}^{\mathrm{ref}}(\eta)$ can be either a ULM or an arbitrary known non-ULM trajectory, and $\Delta \mathbf{p}(\eta)=[\Delta x(\eta),\Delta y(\eta),\Delta z(\eta)]^{\mathrm{T}}$ denotes the unknown position perturbations. We focus on translational position perturbations and assume that the attitude matrix $\mathbf{R}(\eta)$ and the array geometry in the body-fixed coordinate system are known.  
According to (\ref{II-A-eq:position}), the reference position of a Tx/Rx antenna element is $\mathbf{p}_{*}^{\mathrm{ref}}(\eta)=\mathbf{p}^{\mathrm{ref}}(\eta) + \mathbf{R}(\eta)\mathbf{d}_{*}$, and its actual position satisfies
\vspace{-1mm}
\begin{equation}
	\mathbf{p}_{*}^{\mathrm{real}}(\eta) = \mathbf{p}_{*}^{\mathrm{ref}}(\eta) + \Delta\mathbf{p}(\eta), \quad * \in \{p, q\}.
	\label{II-C-eq:Tx/Rx_real_ref_trajectory}
\end{equation}
To facilitate the analysis of the impact of non-ideal motion on conventional RDA, the spatial offsets of the Tx and Sensing Rx elements relative to the UAV platform center are temporarily neglected. Accordingly, the antenna indices $p$ and $q$ are omitted, and the UAV is approximated as a single equivalent phase center following the platform center trajectory.

\textit{1) Known Non-ULM Reference Trajectory}: We first analyze the impact of a known reference trajectory deviating from ULM on conventional RDA. For the $m$-th pixel and each trajectory under consideration, let $\eta_{m,0}$ denote the zero-Doppler slow time, at which the one-way slant range attains its minimum $r_{m,0}$. As a baseline, consider the ideal ULM case where the UAV moves along a straight line at a constant speed $v$. Under the equivalent phase-center approximation, the corresponding two-way slant range can be approximated by its second-order Taylor expansion around $\eta=\eta_{m,0}$ as
\vspace{-1mm}
\begin{equation}
	R_{m}^{\mathrm{ULM}}(\eta)\approx 2r_{m,0} +\frac{v^{2}}{r_{m,0}}(\eta-\eta_{m,0})^{2}.
	\label{II-C-eq:ulm_range}
\end{equation}
In contrast, for an arbitrary known non-ULM reference trajectory $\mathbf{p}^{\mathrm{ref}}(\eta)$, the two-way slant range is $R_{m}^{\mathrm{ref}}(\eta) = 2\|\mathbf p^{\mathrm{ref}}(\eta)-\mathbf t_m\|_2 $. Expanding it around $\eta_{m,0}$ yields
\vspace{-1mm}
\begin{equation}
	\begin{aligned}
		R_m^{\mathrm{ref}}(\eta) \approx  2r_{m,0} + &\Big[ \hat{\mathbf{r}}_{m,0}^{\mathrm{T}} \mathbf{a}^{\mathrm{ref}}(\eta_{m,0})  +\frac{\|\mathbf{v}^{\mathrm{ref}}(\eta_{m,0})\|_2^2}{r_{m,0}} \Big] \\
		& \times (\eta \! -\!  \eta_{m,0})^2 + \mathcal{O}\big((\eta - \eta_{m,0})^3\big),
	\end{aligned}
	\label{II-C-eq:nonulm_range}
\end{equation}
where $ \hat{\mathbf{r}}_{m,0}=\big(\mathbf p^{\mathrm{ref}}(\eta_{m,0}\big)-\mathbf t_m)/r_{m,0}$ is the unit line-of-sight vector to $m$-th pixel, while $\mathbf v^{\mathrm{ref}}(\eta)=\mathrm d\mathbf p^{\mathrm{ref}}(\eta)/\mathrm d\eta$ and $\mathbf a^{\mathrm{ref}}(\eta)=\mathrm d^2\mathbf p^{\mathrm{ref}}(\eta)/\mathrm d\eta^2$ denote the instantaneous velocity and acceleration, respectively.
The term $\mathcal{O}\big((\eta - \eta_{m,0})^3\big)$ collects the third- and higher-order range variations. 

After range processing of the frequency-domain echoes in (\ref{II-B-eq:RF_echo_model}), conventional RDA performs range cell migration correction (RCMC) based on the ULM range model. As indicated by (\ref{II-C-eq:ulm_range}), under ideal ULM, different azimuth pixels within the same range cell exhibit approximately identical second-order range curvature $v^2/r_{m,0}$, allowing a common RCMC mapping to be applied. In contrast, a non-ULM trajectory in (\ref{II-C-eq:nonulm_range}) introduces the pixel-dependent acceleration projection $\hat{\mathbf r}_{m,0}^{\mathrm T}\mathbf a^{\mathrm{ref}}(\eta_{m,0})$ and higher-order range variations. Consequently, even pixels within the same range cell may follow different range-migration trajectories, e.g., the $a$-th and $b$-th pixels in Fig. \ref{fig:system_model}, leading to extreme space variance. Therefore, a ULM-based common RCMC cannot accurately compensate all pixels simultaneously, resulting in residual range errors and range-azimuth coupling.

Beyond RCMC, non-ULM trajectories also impact azimuth focusing. Differentiating the propagation phase term in (\ref{II-B-eq:RF_echo_model}) with respect to $\eta$, the ULM Doppler frequency exhibits a linear variation, enabling RDA to apply a unified azimuth matched filter. Conversely, for non-ULM trajectories
\vspace{-1mm}
\begin{equation}
	\begin{aligned}
		f_{\mathrm{D},m}^{\mathrm{ref}}(k,\eta) \approx &- \frac{2(f_{\mathrm{c}} + k\Delta f)}{c} \Big[ \hat{\mathbf{r}}_{m,0}^{\mathrm{T}} \mathbf{a}^{\mathrm{ref}}(\eta_{m,0}) + \\
		& \frac{\|\mathbf{v}^{\mathrm{ref}}(\eta_{m,0})\|_2^2}{r_{m,0}} \Big] (\eta - \eta_{m,0}) + \mathcal{O}\big((\eta - \eta_{m,0})^2\big),
		\label{eq:nonulm_doppler}
	\end{aligned}
\end{equation}
where the pixel-dependent acceleration projection and higher-order range variations lead to nonlinear and spatial-variant Doppler trajectories, making a common matched filter unable to simultaneously focus different pixels and resulting in azimuth defocusing and geometric distortion.
Moreover, the factor $(f_{\mathrm{c}} + k\Delta f)$ introduces subcarrier-dependent Doppler scaling, broadening the overall Doppler support and thereby increasing the equivalent PRF $f_{\mathrm{PRF}}^{\mathrm{eq}}$ required to avoid slow-time aliasing.

Overall, the main difficulty of conventional SAR algorithms lies in the inadequacy of their decoupling approximation based on the ULM. 
Since the reference trajectory is available, the complete pixel-dependent phase history can be incorporated into an exact measurement operator, thereby avoiding the above model mismatch, as developed in Section III.


$\textit{2) Unknown Position Errors}$: Beyond the known reference trajectory, the actual UAV flight is inevitably affected by unknown position perturbations $\Delta \mathbf{p}(\eta)$ induced by navigation inaccuracies, atmospheric turbulence, and rotor vibrations \cite{moco1}. Unlike the known $\mathbf{p}^{\mathrm{ref}}(\eta)$, these unpredictable errors cause the presumed range and Doppler histories to deviate from actual physical propagation.  
For traditional RDA, such trajectory mismatch directly leads to inaccurate RCMC interpolations and azimuth matched filtering, resulting in severe image defocusing, azimuth smearing, and geometric distortions. Moreover, since $\Delta\mathbf{p}(\eta)$ varies along slow time, a single global calibration cannot adequately compensate for the resulting phase errors.
To fundamentally tackle this unknown non-ideal factor, a comprehensive solution will be detailed in Section IV.

\section{DPS-Based Super-Resolution SAR Imaging}

In this section, we propose a DPS-based super-resolution SAR imaging framework under any known reference trajectory. We first formulate the imaging task as an exact linear inverse problem and cast this ill-posed problem into a Bayesian inference framework. Subsequently, a pre-trained diffusion model serves as a prior for high-fidelity reconstruction, further accelerated by the DDIM sampling mechanism. Throughout this section, the UAV is assumed to follow the known reference trajectory, i.e., $\Delta\mathbf p(\eta)=\mathbf 0$, while unknown position perturbations are addressed in Section IV.

\subsection{Exact Linear Inverse Formulation}
We first address the spatial-variant defocusing induced by the range-azimuth decoupling approximations adopted by conventional SAR algorithms under arbitrary trajectories.
By incorporating the exact Tx–target–Rx propagation geometry associated with the known reference trajectory into the measurement matrix, we establish an exact linear inverse formulation for the SAR imaging problem.
For a complete synthetic aperture, the ISAC echoes in (\ref{II-B-eq:RF_echo_model}) can be stacked over the $K$ subcarriers, $N_{\mathrm{sa}}$ slow-time samples, and $N_{\mathrm{r}}$ Sensing Rx elements, forming $\mathbf{y}\in\mathbb{C}^{L\times 1}$ with $L=K  N_{\mathrm{sa}} N_{\mathrm{r}}$.
Accordingly, the exact measurement matrix $\boldsymbol{\Phi}^{\mathrm{ref}}=[\boldsymbol{\phi}^{\mathrm{ref}}_1,\boldsymbol{\phi}^{\mathrm{ref}}_2,\dots,\boldsymbol{\phi}^{\mathrm{ref}}_{M}] \in  \mathbb{C}^{L\times M}$ is constructed, where the $m$-th column $\boldsymbol{\phi}^{\mathrm{ref}}_m$ acts as the exact spatial response of the $m$-th pixel. Consistent with the stacking order of $\mathbf{y}$, the $(k,\ell,q)$-indexed element of $\boldsymbol{\phi}^{\mathrm{ref}}_m$ is 
\begin{equation}
	\left[\boldsymbol{\phi}^{\mathrm{ref}}_m\right]_{(k,\ell,q)}
	\! =\! \sum_{p=1}^{N_t}A_m^{p,q}[k,\ell]d^p_{u_\ell,k}
	e^{-j2\pi(f_\mathrm{c}+k\Delta f)\tau_m^{p,q}(\eta_\ell)},
	\label{III-A-eq:Phi_element}
\end{equation}
where $\tau_m^{p,q}(\eta_\ell)$ is evaluated using the reference element positions $\mathbf{p}_p^{\rm ref}(\eta_\ell)$ and $\mathbf{p}_q^{\rm ref}(\eta_\ell)$ in (\ref{II-C-eq:Tx/Rx_real_ref_trajectory}). Thus, the complete pixel-dependent propagation phase is retained without trajectory-dependent Taylor approximations or range-azimuth decoupling.
Stacking the noise samples $w^q[k,\ell]$ in the same order as the echoes yields the exact complex-valued linear inverse model
\begin{equation}
	\mathbf{y}= \boldsymbol{\Phi}^{\mathrm{ref}}\boldsymbol{\chi}+\mathbf{w},
	\label{III-A-eq:stacked_complex}
\end{equation}
where $\mathbf{w} \in \mathbb{C}^{L \times 1}$ denotes the stacked frequency domain noise vector satisfying independent and identically distributed complex Gaussian distribution $\mathbf{w} \sim \mathcal{CN}(\mathbf{0},\xi^2\mathbf{I}_L)$ with $\xi^2$ representing the noise variance.
To facilitate subsequent algorithm design, we consider the equivalent real-valued form of (\ref{III-A-eq:stacked_complex}),
\begin{equation}
	\tilde{\mathbf{y}} = \widetilde{\boldsymbol{\Phi}}^{\mathrm{ref}}\mathbf{x}_0 + \tilde{\mathbf{w}},
	\label{III-A-eq:real_value_model}
\end{equation}
where the real-valued observation $\tilde{\mathbf{y}} \in \mathbb{R}^{2L \times 1}$, measurement matrix $\widetilde{\boldsymbol{\Phi}}^{\mathrm{ref}} \in \mathbb{R}^{2L \times 2M}$, and the equivalent reflectivity vector $\mathbf{x}_0 \in \mathbb{R}^{2M \times 1}$ are 
\begin{equation}
	\tilde{\mathbf{y}}\! = \! \begin{bmatrix} \Re\{\mathbf{y}\} \\ \Im\{\mathbf{y}\} \end{bmatrix},  
	\widetilde{\boldsymbol{\Phi}}^{\mathrm{ref}}\!\! =\! \begin{bmatrix} \Re\{\boldsymbol{\Phi}^{\mathrm{ref}}\} & -\Im\{\boldsymbol{\Phi}^{\mathrm{ref}}\} \\ \Im\{\boldsymbol{\Phi}^{\mathrm{ref}}\} & \Re\{\boldsymbol{\Phi}^{\mathrm{ref}}\} \end{bmatrix}, 
	\mathbf{x}_0 = \begin{bmatrix} \Re\{\boldsymbol{\chi}\} \\ \Im\{\boldsymbol{\chi}\} \end{bmatrix}.
\end{equation}
Accordingly, the equivalent real-valued noise follows $\tilde{\mathbf{w}} \sim \mathcal{N}(\mathbf{0}, \xi^2\mathbf{I}_{2L}/2)$.

Although (\ref{III-A-eq:real_value_model}) accurately characterizes the physical observation process, reconstructing the high-dimensional real-valued reflectivity vector $\mathbf{x}_0$ from the observation $\tilde{\mathbf{y}}$ remains a challenging ill-posed inverse problem. 
In practical low-altitude ISAC systems, communication scheduling, target occlusions, and limited slow-time sampling may substantially reduce the number of effective observations. Moreover, adopting a fine imaging grid for super-resolution increases the coherence among the columns of $\widetilde{\boldsymbol{\Phi}}^{\mathrm{ref}}$. Consequently, the inverse problem becomes underdetermined or severely ill-conditioned.
Direct matrix inversion or matched filtering like BP are computationally intractable and sensitive to noise, injecting robust prior knowledge to regularize the solution becomes a necessity.

\vspace{-1mm}
\subsection{Bayesian Inference with Diffusion Prior}
To regularize the severely ill-posed super-resolution inverse problem in (\ref{III-A-eq:real_value_model}), we cast the SAR reconstruction task into a Bayesian inference framework.
Specifically, the objective is to estimate the real-valued reflectivity vector $\mathbf{x}_0$ from the posterior distribution
\vspace{-1mm}
\begin{equation} 
	p(\mathbf{x}_0|\tilde{\mathbf{y}}) \propto p(\tilde{\mathbf{y}}|\mathbf{x}_0) p_0(\mathbf{x}_0), \label{III-B-eq:posterior} 
\end{equation}
where $p(\tilde{\mathbf{y}}|\mathbf{x}_0)$ denotes the likelihood distribution characterizing the physical observation constraint, strictly following
\vspace{-1mm}
\begin{equation}
	p(\tilde{\mathbf{y}}|\mathbf{x}_0) = \mathcal{N}\left(\tilde{\mathbf{y}};\widetilde{\boldsymbol{\Phi}}^{\mathrm{ref}}\mathbf{x}_0,\frac{\xi^2}{2}\mathbf{I}_{2L}\right).
	\label{III-B-eq:likelihood}
\end{equation}
The critical challenge in solving (\ref{III-B-eq:posterior}) lies in constructing a highly expressive prior $p_0(\mathbf{x}_0)$ for SAR reflectivity representations. Conventional SAR reconstruction methods, like compressed sensing algorithms, employ hand-crafted priors, e.g., the $l_1$ sparsity in image or wavelet domain \cite{CS_SAR3}. However, these simplified priors are insufficient to  capture the non-Gaussian statistics, spatial correlations, and structural details of SAR reflectivity images, especially in severely ill-posed or super-resolution settings.
Supervised echo-to-image networks can achieve improved reconstruction performance, but their learned mappings are generally coupled with the measurement operators used during training and may require retraining when the sensing trajectory, MIMO configuration, or OFDM sampling strategy changes.
To overcome these limitations, we employ a diffusion model to learn the prior distribution $p_0(\mathbf{x}_0)$ directly from SAR reflectivity data without any specific structures assumption. Since the diffusion prior is learned independently of the physical forward operator, it can be combined with different trajectories and sensing configurations for posterior inference without retraining.

Specifically, diffusion models learn the underlying data distribution through a self-supervised noise perturbation and denoising process \cite{DDPM}. By progressively corrupting clean samples with Gaussian noise and learning the reverse denoising process, diffusion models capture the underlying statistical characteristics of the SAR reflectivity representations.
Therefore, the ISAC SAR imaging framework consists of two stages, offline diffusion prior training and online posterior inference based on the learned prior and echoes. 
During the offline training phase, clean SAR reflectivity representations $\mathbf{x}_0 \sim p_0(\mathbf{x}_0)$ are sampled from the training dataset. Starting from $\mathbf{x}_0$, the forward diffusion process gradually injects Gaussian noise according to a predefined variance schedule $\{\beta_i\}_{i=1}^N$, where $N$ denotes the total number of diffusion timesteps. Specifically, the transition from the ($i-1$)-th to the $i$-th diffusion state is 
\begin{equation}
	q(\mathbf{x}_i|\mathbf{x}_{i-1})\! =\! \mathcal{N} (\mathbf{x}_i;\! \sqrt{1 - \beta_i}\mathbf{x}_{i-1}, \beta_i\mathbf{I}_{2M}),\quad\!\!\! \!\! i \! =\! 1,\! \dots, \! N. 
	\label{III-B-eq:forward_trans_distribution}
\end{equation}
Defining $\alpha_{i} \triangleq 1 - \beta_{i}$ and $\bar{\alpha}_{i} \triangleq \prod_{i'=1}^{i} \alpha_{i'}$, and applying the reparameterization trick, the noisy reflectivity $\mathbf{x}_i$ at an arbitrary timestep $i$ can be directly obtained as
\begin{equation}
	\mathbf{x}_i = \sqrt{\bar{\alpha}_i}\mathbf{x}_0 + \sqrt{1-\bar{\alpha}_i}\boldsymbol{\epsilon}, \qquad \boldsymbol{\epsilon} \sim \mathcal{N}(\mathbf{0},\mathbf{I}_{2M}).
	\label{III-B-eq:reparameterization}
\end{equation}
With a properly designed variance schedule, $\bar{\alpha}_{N} \approx 0$ when $N$ is sufficiently large, meaning the final diffusion state approximately follows $\mathcal{N}(\mathbf{0}, \mathbf{I}_{2M})$.
That is, the forward process $\mathbf{x}_0 \to \dots \to \mathbf{x}_N$ diffuses the clean reflectivity representation into Gaussian noise.


The reverse diffusion process aims to recover samples from the learned prior distribution by gradually removing the injected noise. This process requires the prior score function
$\nabla_{\mathbf{x}_i}\log p_i(\mathbf{x}_i)$, where $p_i(\mathbf{x}_i)$ denotes the marginal distribution of the noisy state $\mathbf{x}_i$ at timestep $i$. The score function points toward higher-density regions of \(p_i(\mathbf{x}_i)\). However, directly evaluating this score function is intractable due to the unknown data distribution $p_0(\mathbf{x}_0)$.  
Based on denoising score matching theory \cite{DSM}, estimating this intractable score is equivalent to matching the tractable conditional score $\nabla_{\mathbf{x}_i} \log q(\mathbf{x}_i|\mathbf{x}_0)$.
Given the forward Gaussian transition in (\ref{III-B-eq:reparameterization}), the conditional score function is
\begin{equation}
	\nabla_{\mathbf{x}_i} \log q(\mathbf{x}_i|\mathbf{x}_0) = -\frac{\mathbf{x}_i - \sqrt{\bar{\alpha}_i} \mathbf{x}_0}{1 - \bar{\alpha}_i} = -\frac{1}{\sqrt{1 - \bar{\alpha}_i}} \boldsymbol{\epsilon}.
\end{equation}
Since the  conditional score is proportional to the injected noise $\boldsymbol{\epsilon}$, we train a parameterized U-Net $\boldsymbol{\epsilon}_{\boldsymbol{\theta}}(\mathbf{x}_i,i)$ to predict the injected noise component $\boldsymbol{\epsilon}$. Consequently, the network parameters are optimized via the mean squared error loss
\begin{equation}
	\mathcal{L}(\boldsymbol{\theta}) = \mathbb{E}_{\mathbf{x}_0, \boldsymbol{\epsilon}, i} \left[ \left\| \boldsymbol{\epsilon} - \boldsymbol{\epsilon}_{\boldsymbol{\theta}}(\sqrt{\bar{\alpha}_i}\mathbf{x}_0 + \sqrt{1 - \bar{\alpha}_i}\boldsymbol{\epsilon}, i) \right\|_2^2 \right].
	\label{eq:loss}
\end{equation}
After convergence, the learned noise predictor provides an approximation of the prior score function as
\begin{equation}
	\nabla_{\mathbf{x}_i} \log p_i(\mathbf{x}_i) \approx - \frac{1}{\sqrt{1 - \bar{\alpha}_i}} \boldsymbol{\epsilon}_{\boldsymbol{\theta}}(\mathbf{x}_i, i). \label{eq:prior_score}
\end{equation}
Operating exclusively on the intermediate $\mathbf{x}_i$ and the sinusoidal positional embedding of step $i$, the optimized network $\boldsymbol{\epsilon}_{\boldsymbol{\theta}}$ provides a highly expressive, purely data-driven prior score via \eqref{eq:prior_score}.
Since this offline learning process strictly involves only SAR reflectivity data, the resulting universal prior remains completely independent of the trajectory dependent measurement matrix $\widetilde{\boldsymbol{\Phi}}^{\mathrm{ref}}$ and observation noise. More importantly, the learned diffusion prior captures rich spatial structures and correlations of SAR reflectivity, providing strong regularization for the severely ill-posed inverse problem induced by fine-grid reconstruction and thereby enabling super-resolution recovery beyond the physical resolution.
The crucial next step is to embed these specific physical observation constraints into the reverse process for online posterior inference.

\begin{figure}
	\centering
	\includegraphics[width=0.49\textwidth]{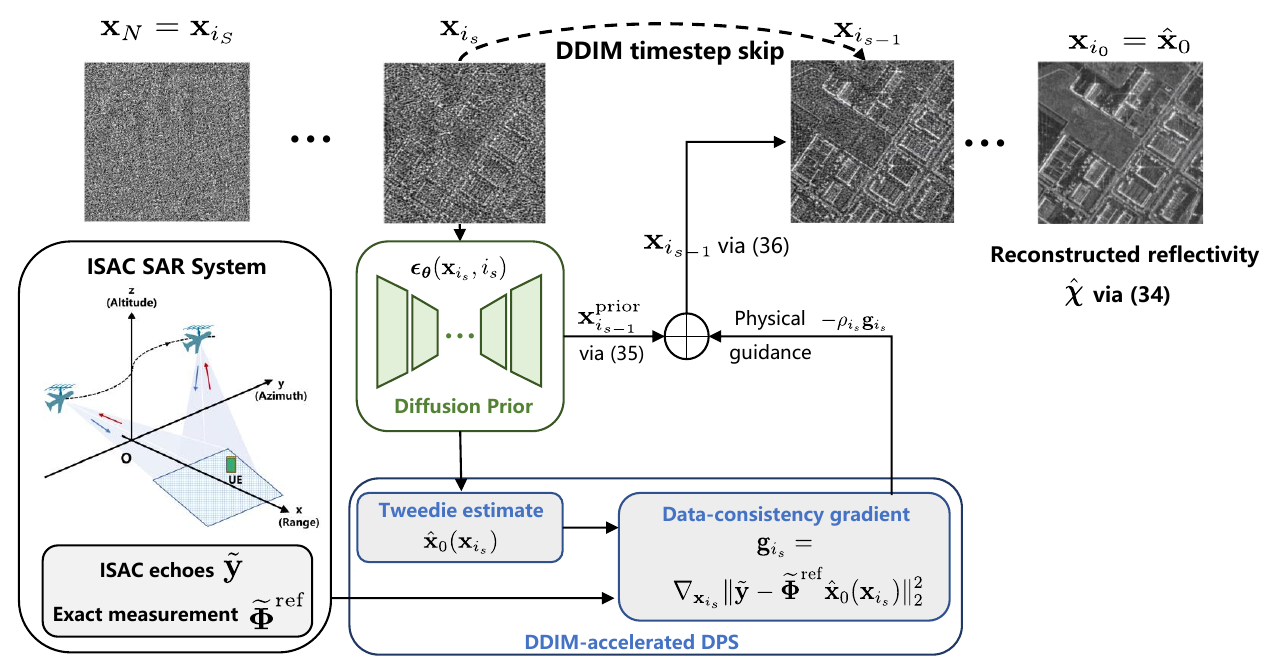}
	\caption{DDIM-accelerated DPS for super-resolution ISAC SAR imaging.}
	\label{fig:DDIM_DPS}
		\vspace{-3mm}
\end{figure}

\subsection{DPS-based SAR Imaging}
Following the offline training of the diffusion prior, we employ DPS to reconstruct the unknown clean reflectivity $\mathbf{x}_0$ by incorporating the observed ISAC echoes $\tilde{\mathbf{y}}$ into the reverse diffusion process. Specifically, at each diffusion timestep $i$, DPS utilizes the posterior score function $\nabla_{\mathbf{x}_i}\log p_i(\mathbf{x}_i|\tilde{\mathbf{y}})$ to guide the reverse diffusion trajectory toward samples consistent with both the learned prior and the physical observations. 
According to Bayes' rule, the posterior score is decomposed as
\begin{equation}
	\nabla_{\mathbf{x}_i} \log p_i(\mathbf{x}_i | \tilde{\mathbf{y}}) = \nabla_{\mathbf{x}_i} \log p_i(\mathbf{x}_i) + \nabla_{\mathbf{x}_i} \log p_i(\tilde{\mathbf{y}} | \mathbf{x}_i),
	\label{eq:posterior_score}
\end{equation}
where the first term is the prior score provided by the pre-trained network $\boldsymbol{\epsilon}_{\boldsymbol{\theta}}$ through \eqref{eq:prior_score}, while the second term is the likelihood score introducing the physical observation constraint into the reverse diffusion process.

The main difficulty lies in evaluating the likelihood score $\nabla_{\mathbf{x}_i}\log p_i(\tilde{\mathbf{y}}|\mathbf{x}_i)$, because the physical likelihood in (\ref{III-B-eq:likelihood}) is defined for the clean reflectivity $\mathbf{x}_0$, whereas $\mathbf{x}_i$ is a noisy intermediate state. 
Using the conditional independence of $\mathbf{x}_i$ and $\tilde{\mathbf{y}}$ given $\mathbf{x}_0$, the likelihood at timestep $i$ can be expressed by marginalizing over $\mathbf{x}_0$  \cite{song2023pseudoinverse} 
\begin{equation}
	p_i(\tilde{\mathbf{y}} | \mathbf{x}_i) = \int p(\tilde{\mathbf{y}} | \mathbf{x}_0) p(\mathbf{x}_0 | \mathbf{x}_i) \mathrm{d}\mathbf{x}_0.
	\label{III-C-eq:marginalization}
\end{equation}
Since the marginalization over $\mathbf{x}_0$ is intractable, DPS approximates (\ref{III-C-eq:marginalization}) by evaluating the likelihood at a representative clean reflectivity  estimate $\hat{\mathbf{x}}_0(\mathbf{x}_i)$ derived from current state $\mathbf{x}_i$ \cite{DPS}. Specifically, the conditional mean $\mathbb{E}[\mathbf{x}_0|\mathbf{x}_i]$ is adopted as the representative estimate of $\hat{\mathbf{x}}_0(\mathbf{x}_i)$.
Combining Tweedie's formula \cite{song2023pseudoinverse} with the learned prior score approximation yields
\begin{equation}
	\hat{\mathbf{x}}_0(\mathbf{x}_i)= \frac{\mathbf{x}_i - \sqrt{1 - \bar{\alpha}_i}\boldsymbol{\epsilon}_{\boldsymbol{\theta}}(\mathbf{x}_i, i)}{\sqrt{\bar{\alpha}_i}} \approx \mathbb{E}[\mathbf{x}_0 | \mathbf{x}_i] .
	\label{eq:tweedie}
\end{equation}
The marginal likelihood is then approximated by evaluating $p(\tilde{\mathbf{y}}|\mathbf{x}_0)$ at $\hat{\mathbf{x}}_0(\mathbf{x}_i)$, i.e.,
\vspace{-1mm}
\begin{equation}
	p_i(\tilde{\mathbf{y}}|\mathbf{x}_i)
	\! \approx \!p(\tilde{\mathbf{y}}|\hat{\mathbf{x}}_0(\mathbf{x}_i))=
	\mathcal{N}
	\Big(
	\tilde{\mathbf{y}};
	\widetilde{\boldsymbol{\Phi}}^{\mathrm{ref}}
	\hat{\mathbf{x}}_0(\mathbf{x}_i),
	\frac{\xi^2}{2}\mathbf{I}_{2L}
	\Big).
	\label{eq:dps_approx}
\end{equation}
Accordingly, the likelihood score can be approximated as
\vspace{-1mm}
\begin{equation}
	\nabla_{\mathbf{x}_i} \log p_i(\tilde{\mathbf{y}} | \mathbf{x}_i) \approx 
	- {\xi^{-2}}\mathbf{g}_i , 
	\label{eq:simplified_likelihood}
\end{equation}
where $\mathbf{g}_i$ is the physical data-consistency gradient, given by
\vspace{-1mm}
\begin{equation}
	\mathbf{g}_i = \nabla_{\mathbf{x}_i} \| \tilde{\mathbf{y}}-\widetilde{\boldsymbol{\Phi}}^{\mathrm{ref}}
	\hat{\mathbf{x}}_0(\mathbf{x}_i)\|_2^2.
	\label{III-C-eq:gradient}
\end{equation} 
Although $\xi^{-2}$ provides the nominal likelihood scaling, the measurement noise variance $\xi^2$ may be unknown or imperfectly estimated in practical systems. Moreover, the magnitude of $\mathbf{g}_i$ can vary substantially across diffusion timesteps. Therefore, rather than directly using the fixed scaling $\xi^{-2}$, we employ an adaptive guidance coefficient
\vspace{-1mm}
\begin{equation}
	\rho_i = \frac{\zeta}{\|\mathbf{g}_i\|_2 + \delta}  \|\boldsymbol{\epsilon}_{\boldsymbol{\theta}}(\mathbf{x}_i, i)\|_2,  \label{III-C-eq:rho_step_size}
\end{equation}
where $\zeta>0$ controls the overall guidance strength and $\delta>0$ is a small constant for numerical stability. The resulting physical guidance is implemented as $-\rho_i\mathbf{g}_i$.

Finally, to realize posterior sampling, DPS incorporates the approximated likelihood score as physical guidance into the standard DDPM ancestral sampling process. Specifically, the DDPM sampler first generates a prior-driven intermediate state
\vspace{-1mm}
\begin{equation}
	\mathbf{x}_{i-1}^{\mathrm{prior}} = \frac{1}{\sqrt{\alpha_i}} \left( \mathbf{x}_i - \frac{1-\alpha_i}{\sqrt{1-\bar{\alpha}_i}} \boldsymbol{\epsilon}_{\boldsymbol{\theta}}(\mathbf{x}_i, i) \right)  + \nu_i \mathbf{z}, \label{eq:dps_prior}
\end{equation}
where  $\mathbf{z} \sim \mathcal{N}(\mathbf{0},\mathbf{I}_{2M})$ represents the injected stochastic noise with scale $\nu_i$. The physical guidance is then applied to enforce consistency with the echoes,
\begin{equation}
	\mathbf{x}_{i-1} =  \mathbf{x}_{i-1}^{\mathrm{prior}}- \rho_i \mathbf{g}_i. \label{eq:dps_update}
\end{equation}
Iteratively sampling from $i=N$ to 1 yields the reconstructed real-valued $\hat{\mathbf{x}}_0$, from which the final complex SAR image $\hat{\boldsymbol{\chi}}$ is
\begin{equation}
\hat{\chi}_m = [\hat{\mathbf{x}}_0]_m + j[\hat{\mathbf{x}}_0]_{M+m}, \qquad m=1,\ldots,M.
	\label{eq:real2complex}
\end{equation}

\begin{algorithm}[t]
	\caption{DDIM-Accelerated DPS for ISAC SAR Imaging}
	\label{alg:dps_ddim}
	\begin{algorithmic}[1]
		\REQUIRE Echoes $\tilde{\mathbf{y}}$, measurement matrix $\widetilde{\boldsymbol{\Phi}}^{\mathrm{ref}}$, pre-trained diffusion prior network $\boldsymbol{\epsilon}_{\boldsymbol{\theta}}$,  DDIM sampling steps $S$, and guidance parameters $\zeta$ and $\delta$. \\ 
		\ENSURE Reconstructed complex SAR reflectivity $\hat{\boldsymbol{\chi}}$.
		\STATE Construct the uniform DDIM sampling timestep subsequence
		$\{i_s\}_{s=0}^{S}$ with
		$i_s=\left\lfloor sN/S \right\rfloor$. 
		\STATE \textbf{Initialize:} Sample from Gaussian noise $\mathbf{x}_{i_S}\sim \mathcal{N}(\mathbf{0},\mathbf{I}_{2M})$.
		\FOR{$s = S$ to 1}
		\STATE Predict $\boldsymbol{\epsilon}_{\boldsymbol{\theta}}(\mathbf{x}_{i_s},i_s)$ using the pre-trained diffusion network.
		\STATE Compute the clean reflectivity estimate $\hat{\mathbf{x}}_0(\mathbf{x}_{i_s})$ via (\ref{eq:tweedie}).
		\STATE Compute the DDIM-predicted prior state $\mathbf{x}_{i_{s-1}}^{\mathrm{prior}}$ via (\ref{eq:ddim_prior}).
		\STATE Compute the physical gradient $\mathbf{g}_{i_s}$ via (\ref{III-C-eq:gradient}) and adaptive coefficient $\rho_{i_s}$ via (\ref{III-C-eq:rho_step_size}).
		\STATE Update $\mathbf{x}_{i_{s-1}}$ via (\ref{eq:ddim_dps}).
		\ENDFOR
		\STATE  Obtain the final real-valued SAR reflectivity $\hat{\mathbf{x}}_0={\mathbf{x}}_{i_0}$.
		\STATE Recover the complex SAR reflectivity $\hat{\boldsymbol{\chi}}$ via (\ref{eq:real2complex}).
	\end{algorithmic}
	
\end{algorithm}


\subsection{DDIM Acceleration}
Although the DPS method in (\ref{eq:dps_update}) enables high-quality reconstruction, the standard DDPM ancestral sampler sequentially traverses the adjacent diffusion timesteps from $i=N$ to $i=1$. Since each reverse step requires an evaluation of the diffusion network, a large $N$  (e.g., $N \ge 1000$ in \cite{DDPM}) introduces prohibitive computational cost and reconstruction latency for low-altitude ISAC applications. We therefore adopt the DDIM sampler \cite{DDIM} to accelerate the reverse process by performing deterministic sampling over a sparse subsequence of the original diffusion timesteps.
Specifically, let $N \! =\! i_S \! > \! i_{S-1} \! > \! \cdots>i_1>i_0=0$ denote the selected timestep subsequence, where $S\ll N$ is the number of DDIM sampling steps and is fixed during reconstruction. Here, $s=1,\ldots,S$ indexes the DDIM sampling steps, whereas $i_s$ denotes the corresponding timestep in the original diffusion process, with $\bar{\alpha}_{i_0}\! = \! 1$.
In this work, the subsequence is uniformly selected as $i_s \! =\! \left\lfloor sN/S\right\rfloor$.

At the current DDIM sampling step $s$, the diffusion network takes $(\mathbf{x}_{i_s},i_s)$ as input, and the clean reflectivity estimate $\hat{\mathbf{x}}_0(\mathbf{x}_{i_s})$ can be obtained using (\ref{eq:tweedie}).
Under the deterministic DDIM setting, the prior-driven state from $i_s$ to $i_{s-1}$ is 
\begin{equation}
	\mathbf{x}_{i_{s-1}}^{\mathrm{prior}}
	=
	\sqrt{\bar{\alpha}_{i_{s-1}}}
	\hat{\mathbf{x}}_0(\mathbf{x}_{i_s})
	+
	\sqrt{1-\bar{\alpha}_{i_{s-1}}}
	\boldsymbol{\epsilon}_{\boldsymbol{\theta}}
	(\mathbf{x}_{i_s},i_s).
	\label{eq:ddim_prior}
\end{equation}
Unlike the adjacent DDPM update, this transition directly maps $i_s$ to the selected timestep $i_{s-1}$ without additional stochastic noise injection.
Following the physical guidance defined in (\ref{III-C-eq:gradient}) and $(\ref{III-C-eq:rho_step_size})$, $\mathbf{g}_{i_s}$ and $\rho_{i_s}$ are evaluated at the current selected timestep $i_s$. The DDIM-accelerated DPS update is 
\begin{equation}
	\mathbf{x}_{i_{s-1}}
	=
	\mathbf{x}_{i_{s-1}}^{\mathrm{prior}}
	-
	\rho_{i_s}\mathbf{g}_{i_s}.
	\label{eq:ddim_dps}
\end{equation}
Repeating the above update for $s=S,\ldots,1$ reduces the number of diffusion network evaluations from $N$ to $S$ while retaining the physical data-consistency guidance at each selected timestep. The complete procedure is summarized in Algorithm \ref{alg:dps_ddim} and illustrated in Fig. \ref{fig:DDIM_DPS}.

\section{Joint Iterative DPS and MoCo Framework}
This section extends the reference-trajectory-based DPS framework to account for unknown UAV position errors by developing a joint DPS and MoCo approach. 
The diffusion-guided SAR reconstruction and position error refinement are alternately coupled through an iterative inference framework to mitigate the operator mismatch induced by flight perturbations. 
Furthermore, a DCT-based error reparameterization and a coarse-to-fine refinement strategy are introduced to improve estimation stability and efficiency.

	\vspace{-2mm}
\subsection{Joint Posterior Inference under Unknown Position Errors}
The exact inverse problem formulation and DDIM-accelerated DPS developed in Section III enable super-resolution SAR imaging under arbitrary known reference trajectories $\mathbf{p}_{*}^{\mathrm{ref}}(\eta)$. 
In practical low-altitude ISAC systems, however, unknown Tx/Rx position errors $\Delta\mathbf{p}(\eta)$ perturb the actual UAV trajectories and lead to measurement operator mismatch.
Collecting the position errors over the synthetic aperture as $\mathbf{E}=[\Delta\mathbf{p}(\eta_0),\dots,\Delta\mathbf{p}(\eta_{N_{\mathrm{sa}}-1})]\in\mathbb{R}^{3\times N_{\mathrm{sa}}}$,
the corresponding inverse problem with actual antenna positions in (\ref{II-C-eq:Tx/Rx_real_ref_trajectory}) yields 
\begin{equation}
	\tilde{\mathbf{y}} = \widetilde{\boldsymbol{\Phi}}(\mathbf{E})\mathbf{x}_0 + \tilde{\mathbf{w}},
	\label{eq:moco_inverse}
\end{equation}
where $\widetilde{\boldsymbol{\Phi}}(\mathbf{E})$ denotes the measurement operator constructed from the position-error-corrupted Tx/Rx trajectories and reduces to $\widetilde{\boldsymbol{\Phi}}^{\mathrm{ref}}$ when $\mathbf{E}=\mathbf{0}$. Hence, unknown position errors invalidate the reference operator $\widetilde{\boldsymbol{\Phi}}^{\mathrm{ref}}$ in Section III and corrupt the DPS data-consistency guidance  $\mathbf{g}_{i_s}$, degrading coherent SAR reconstruction. To mitigate this mismatch, we develop a joint DPS and MoCo framework that couples reflectivity reconstruction with iterative position error refinement throughout the reverse diffusion process.

Conventional MoCo or autofocus methods typically rely on isolated dominant scatterers and assume spatially invariant phase errors under the ULM approximation \cite{PGA}. However, the flexible trajectories of the MIMO-OFDM spotlight mode induce severe slow-time- and space-variant couplings. Consequently, traditional approximations fail, particularly in complex scenes lacking ideal point scatterers.
We therefore formulate the SAR reflectivity reconstruction and position error estimation within a joint posterior inference framework as
\begin{equation}
	p(\mathbf{x}_0, \mathbf{E} | \tilde{\mathbf{y}}) \propto p(\tilde{\mathbf{y}} | \mathbf{x}_0, \mathbf{E}) p_0(\mathbf{x}_0) p(\mathbf{E}), 
	\label{joint_posterior}
\end{equation}
where $\mathbf{x}_0$ and $\mathbf{E}$ are assumed to be a priori independent. The reflectivity prior $p_0(\mathbf{x}_0)$ is characterized by the pre-trained diffusion model, whereas $p(\mathbf{E})$ is the structural prior on position errors.
Directly extending DPS to jointly sample $\mathbf{x}_0$ and $\mathbf{E}$
from (\ref{joint_posterior}) is impractical. Unlike the reflectivity prior which can be provided by the diffusion model, $p(\mathbf{E})$ is highly intractable to explicitly formulate or directly learn via deep learning methods. Moreover, the non-linear dependence of $\widetilde{\boldsymbol{\Phi}}(\mathbf{E})$
on $\mathbf{E}$ results in strong coupling between the reflectivity and position errors, making direct high-dimensional joint sampling difficult and numerically unstable.

To address this difficulty, we develop an alternating inference scheme that embeds iterative position error refinement into the DDIM-accelerated DPS reverse process. Specifically, at the $s$-th DDIM sampling step, the current noisy reflectivity $\mathbf{x}_{i_s}$ is first mapped to the clean reflectivity estimate $\hat{\mathbf{x}}_0(\mathbf{x}_{i_s})$ using (\ref{eq:tweedie}).
Conditioned on this estimate, the position error is then refined through
$N_{\mathrm{MoCo}}$ MoCo iterations. Let $\mathbf{E}_{s,n}$ denote the position-error estimate at the
$n$-th MoCo iteration of DDIM sampling step $s$, where $n=0,\ldots,N_{\mathrm{MoCo}}-1$.
The corresponding data consistency loss is
\begin{equation}
	\mathcal{L}_{s}(\mathbf{E}_{s,n})
	=\| \tilde{\mathbf{y}}	- \widetilde{\boldsymbol{\Phi}}(\mathbf{E}_{s,n})
	\hat{\mathbf{x}}_0(\mathbf{x}_{i_s})\|_2^2.
	\label{joint_loss}
\end{equation}
The time-domain position-error update can be expressed as
\begin{equation}
	\mathbf{E}_{s,n+1} = \mathbf{E}_{s,n} - \gamma \nabla_{\mathbf{E}_{s,n}} \mathcal{L}_{s}(\mathbf{E}_{s,n}), 
\end{equation}
where $\gamma>0$ is the position-error learning rate.
For the first DDIM sampling step, the position-error estimate is initialized as $\mathbf{E}_{S,0}=\mathbf{0}$. The refined estimate from each sampling step is used to initialize the subsequent step, i.e., $\mathbf{E}_{s-1,0}=\mathbf{E}_{s,N_{\mathrm{MoCo}}}$.

After $N_{\mathrm{MoCo}}$ MoCo refinements, the updated measurement operator
$\widetilde{\boldsymbol{\Phi}}(\mathbf{E}_{s,N_{\mathrm{MoCo}}})$ is used to evaluate  $\mathbf{g}_{i_s}$ for the DDIM-accelerated DPS update from $\mathbf{x}_{i_s}$ to $\mathbf{x}_{i_{s-1}}$. Consequently, progressively improved reflectivity estimates facilitate position-error refinement, while the refined position errors provide more accurate physical guidance for subsequent DPS updates.

\begin{algorithm}[t]
	\caption{Joint Iterative DPS and MoCo}
	\label{alg:joint_moco}
	\begin{algorithmic}[1]
		\REQUIRE Echoes $\tilde{\mathbf{y}}$, reference trajectories $\mathbf{p}_{*}^{\mathrm{ref}}(\eta)$, pre-trained diffusion network $\boldsymbol{\epsilon}_{\boldsymbol{\theta}}$, DDIM sampling steps $S$, MoCo refinements $N_{\mathrm{MoCo}}$, learning rate $\gamma$, cutoff-frequency schedule $f_{\mathrm{cut}}(s)$, and guidance parameters $\zeta$ and $\delta$.
		\ENSURE Reconstructed complex SAR reflectivity  $\hat{\boldsymbol{\chi}}$ and estimated position error $\hat{\mathbf{E}}$.
		
		\STATE Construct uniform DDIM timestep subsequence $\{i_s\}_{s=0}^{S}$.
		
		\STATE \textbf{Initialize:} $\mathbf{x}_{i_S}\sim\mathcal{N}(\mathbf{0},\mathbf{I}_{2M})$
		and $\mathbf{E}_{S,0}=\mathbf{0}$.
		
		\FOR{$s = S$ \TO $1$}

		\STATE Predict $\boldsymbol{\epsilon}_{\boldsymbol{\theta}}(\mathbf{x}_{i_s},i_s)$ and obtain $\hat{\mathbf{x}}_0(\mathbf{x}_{i_s})$ via (\ref{eq:tweedie}).
		\STATE Determine $N_{\mathrm{DCT},s}$ from $f_{\mathrm{cut}}(s)$ via (\ref{eq:active_mode_number}) and initialize $\mathbf{C}_{s,0}$ via (\ref{eq:dct_init}).
		
		\STATE // \textit{--- Position-Error MoCo Refinement ---}
		\FOR{$n=0$ \TO $N_{\mathrm{MoCo}}-1$}
		\STATE Reconstruct $\mathbf{E}_{s,n}$ from $\mathbf{C}_{s,n}$ via (\ref{eq:dct_reconstruction}).
		\STATE Construct $\widetilde{\boldsymbol{\Phi}}(\mathbf{E}_{s,n})$
		and update $\mathbf{C}_{s,n+1}$ via (\ref{eq:act_update}).
		\ENDFOR
		
		\STATE Reconstruct the refined position error
		$\mathbf{E}_{s,N_{\mathrm{MoCo}}}$
		via (\ref{eq:dct_reconstruction}).
		
		\STATE // \textit{--- DDIM-DPS Update ---}
		\STATE Compute the DDIM-predicted prior state
		$\mathbf{x}_{i_{s-1}}^{\mathrm{prior}}$
		via (\ref{eq:ddim_prior}).
		
		\STATE Compute
		$\mathbf{g}_{i_s}$ using $ \widetilde{\boldsymbol{\Phi}}(\mathbf{E}_{s,N_{\mathrm{MoCo}}})$ via (\ref{III-C-eq:gradient}) and $\rho_{i_s}$ via (\ref{III-C-eq:rho_step_size}).
		
		\STATE Update $\mathbf{x}_{i_{s-1}}$
		via (\ref{eq:ddim_dps}).
		
		\IF{$s>1$}
		\STATE Set $\mathbf{E}_{s-1,0}\leftarrow\mathbf{E}_{s,N_{\mathrm{MoCo}}}$ for the next DDIM sampling.
		\ENDIF
		\ENDFOR
		
		\STATE Obtain the final estimate $\hat{\mathbf{x}}_0=\mathbf{x}_{i_0}$ and $\hat{\mathbf{E}}=
		\mathbf{E}_{1,N_{\mathrm{MoCo}}}$.
		\STATE Recover the complex SAR reflectivity $\hat{\boldsymbol{\chi}}$ via (\ref{eq:real2complex}).
	\end{algorithmic}
\end{algorithm}

	\vspace{-3mm}
\subsection{Efficient Position Error Optimization}
Although the alternating DPS and MoCo framework enables joint SAR reconstruction and position-error refinement, directly optimizing the time-domain position-error matrix
$\mathbf{E}_{s,n}\in\mathbb{R}^{3\times N_{\mathrm{sa}}}$ involves $3N_{\mathrm{sa}}$ optimization variables and is highly sensitive to unreliable intermediate reflectivity estimates.
Specifically, independent updates over slow-time samples may introduce rapid and physically implausible fluctuations into the estimated position errors. To reduce the optimization dimension and improve the robustness of MoCo, we reparameterize the position errors in DCT domain and
develop a coarse-to-fine spectral scheduling strategy within the reverse diffusion process.

\textsl{1) DCT Reparameterization}: 
UAV position errors generally exhibit strong temporal correlation along slow time, such that large-scale deviations can be represented by a limited number of low-order DCT modes.
In contrast, observation noise and inaccurate intermediate reflectivity estimates may induce unsupported high-frequency variations. This motivates restricting the MoCo refinement to a low-dimensional DCT subspace.
At the $s$-th DDIM sampling step, let $N_{\mathrm{DCT},s}$ denote the number of active DCT modes determined by the cutoff frequency $f_{\mathrm{cut}}(s)$, i.e.,
\begin{equation}
	N_{\mathrm{DCT},s} = |\{\kappa\in\{0,\ldots,N_{\mathrm{sa}}-1\}:f_{\kappa}\leq f_{\mathrm{cut}}(s)\}|,
	\label{eq:active_mode_number}
\end{equation}
where $f_{\kappa}$ denotes the equivalent temporal frequency of the $\kappa$-th DCT mode. Since the DCT modes are ordered from low to high temporal frequencies, the first $N_{\mathrm{DCT},s}$ modes are retained for position-error refinement.
The active DCT coefficients are initialized from the current position-error estimate as
\begin{equation}
	[\mathbf{C}_{s,0}]_{a,\kappa}
\!	= \!
	[\mathrm{DCT}(\mathbf{E}_{s,0})]_{a,\kappa}, 
	\quad \!
	\begin{aligned}
		a &= 1,2,3,\\
		\kappa &= 0,\ldots,N_{\mathrm{DCT},s}-1,
	\end{aligned}
	\label{eq:dct_init}
\end{equation}
where $\mathbf{C}_{s,0}\!\in\! \mathbb{R}^{3\times N_{\mathrm{DCT},s}}$ collects the active low-frequency DCT coefficients. During the $n$-th MoCo refinement, the full length position-error is reconstructed by zero-padding the inactive DCT modes and performing an $N_{\mathrm{sa}}$-point inverse DCT,
\begin{equation}
	\mathbf{E}_{s,n}
	=
	\mathrm{IDCT}
	\left(
	\left[
	\mathbf{C}_{s,n},
	\mathbf{0}_{3\times(N_{\mathrm{sa}}-N_{\mathrm{DCT},s})}
	\right]
	\right)
	\in\mathbb{R}^{3\times N_{\mathrm{sa}}}.
	\label{eq:dct_reconstruction}
\end{equation}
The reconstructed $\mathbf{E}_{s,n}$ is then used to construct $\widetilde{\boldsymbol{\Phi}}(\mathbf{E}_{s,n})$ and evaluate the data-consistency loss in (\ref{joint_loss}).
The active DCT coefficients are updated according to
\begin{equation}
	\mathbf{C}_{s,n+1}
	=
	\mathbf{C}_{s,n}
	-
	\gamma
	\nabla_{\mathbf{C}_{s,n}}
	\mathcal{L}_{s}(\mathbf{E}_{s,n}),
	\label{eq:act_update}
\end{equation}
where the gradient is backpropagated through the zero-padding, IDCT, and
position-error-dependent measurement operator.
Compared with directly optimizing $\mathbf{E}_{s,n}\in\mathbb{R}^{3\times N_{\mathrm{sa}}}$, the proposed DCT reparameterization reduces the number of optimized variables from $3N_{\mathrm{sa}}$ to $3N_{\mathrm{DCT},s}$. This reduced search space suppresses unsupported high-frequency fluctuations and improves the  efficiency of the MoCo refinement.

\textsl{2) Coarse-to-fine (C2F) Scheduling}: 
The reliability of the clean reflectivity estimate $\hat{\mathbf{x}}_0(\mathbf{x}_{i_s})$ improves progressively during the reverse diffusion process. At early DDIM sampling steps, where $s$ is close to $S$, the reflectivity estimate remains relatively noisy and is therefore insufficient for reliably recovering fine-scale position-error variations.
Accordingly, we adopt a small cutoff frequency $f_{\mathrm{cut}}(s)$, such that only a limited number $N_{\mathrm{DCT},s}$ of low-order DCT modes are activated for MoCo refinement.
As the reverse diffusion proceeds from $s=S$ to $s=1$, $f_{\mathrm{cut}}(s)$ is gradually increased to expand the active DCT subspace, yielding $N_{\mathrm{DCT},S} \leq N_{\mathrm{DCT},S-1} \leq \cdots \leq
N_{\mathrm{DCT},1}$. This C2F expansion first captures dominant low-frequency position errors and progressively incorporates finer variations as the reflectivity estimate improves, thereby enhancing the robustness of MoCo refinement and the physical data-consistency guidance for DPS reconstruction.
The joint DPS and MoCo process is summarized in Algorithm \ref{alg:joint_moco}.

\begin{table}[tbp]
	\centering
	\caption{Main Simulation Parameters of the Proposed Framework}
	\label{tab:parameters}
	\renewcommand{\arraystretch}{1} 
	\begin{tabular}{@{}llc@{}}
		\toprule
		\textbf{Category} & \textbf{Parameters} & \textbf{Value} \\ 
		\midrule
		\multirow{7}{1.8cm}{\textbf{Diffusion-Prior Training}} 
		& Base channels  & $128$\\
		& Multipliers & $[1, 1, 2, 2, 4, 4]$ \\
		& Training timesteps ($N$) & $1000$ \\
		& Variance schedule ($\beta_1 \sim \beta_N$) & Linear, $10^{-4} \sim 0.02$ \\
		& Epochs & $100$ \\
		& Learning rate & $5 \times 10^{-5}$\\
		& Scheduler &  Cosine \\ 
		\midrule
		\multirow{3}{1.8cm}{\textbf{DDIM-DPS}} 
		& Accelerated steps ($S$) & $20$ \\
		& Guidance parameter ($\zeta$) & $2$ \\ 
		& Numerical constant ($\delta$) & $10^{-8}$ \\
		\midrule
		\multirow{3}{1.8cm}{\textbf{DPS $\&$ MoCo}} 
		& Motion learning rate ($\gamma$) & $0.01$ \\
		& Inner iterations ($N_{\mathrm{MoCo}}$) & $15$ \\
		& Active DCT modes($N_{\mathrm{DCT},s}$) & $3 \rightarrow 128$ \\
		\bottomrule
	\end{tabular}
		\vspace{-3mm}
\end{table}

\section{Numerical Results}
\subsection{Simulation Setup}
We consider a low-altitude UAV equipped with an ISAC SAR system and operating as a dual-functional aerial BS. The UAV simultaneously serves ground communication users and performs spotlight SAR imaging using the echoes of the downlink OFDM communication signals. Its initial altitude and speed are set to $H_{\mathrm{init}}=150$ m and $v_{\mathrm{init}}=15$ m/s, respectively, while the specific flight trajectory is configured according to each experimental scenario. 
The ground communication users and the target scene are located within a 100 m $\times$ 100 m area, which remains illuminated throughout the synthetic aperture. The UAV is equipped with the MIMO ULAs of $N_{\mathrm{t}}=4$ ISAC Tx and $N_{\mathrm{r}}=4$ Sensing Rx, and it employs the OFDM waveform with a bandwidth $B=400$ MHz at $f_{\mathrm{c}}=4.9$ GHz. Considering the antenna beamwidth, look angle, target-area extent, and instantaneous flight geometry, the minimum and maximum total propagation distances are set to $R_{\mathrm{near}}=360$ m and $R_{\mathrm{far}}=480$ m, respectively. The principal OFDM and slow-time sampling parameters are selected according to the following sensing and communication constraints:
\begin{itemize}
	\item \textbf{Subcarrier Spacing} $\Delta f$: To avoid ambiguity over the maximum propagation delay, the subcarrier spacing must satisfy $\Delta f \leq c /  R_{\mathrm{far}} \approx 625$ kHz. We set $\Delta f = 120$ kHz to balance the communication symbol duration.
	\item \textbf{CP Length} $T_{\mathrm{cp}}$: To cover the propagation-delay spread across the imaging swath and avoid inter-range-cell interference, the CP is required with $T_{\mathrm{cp}} \ge (R_{\mathrm{far}}-R_{\mathrm{near}})/c$ \cite{OFDM_CP}. We set $T_{\mathrm{cp}}=0.2T_{\mathrm{d}}$, which satisfies the requirement while maintaining a moderate communication overhead.
	\item \textbf{Equivalent slow-time PRF $f_{\mathrm{PRF}}^{\mathrm{eq}}$}:
	 To prevent azimuth aliasing, $f_{\mathrm{PRF}}^{\mathrm{eq}}$ should exceed the maximum Doppler bandwidth $B_{\mathrm{D}}^{\mathrm{max}}$ \cite{SAR_survey}. Following (\ref{II-B-eq:PRF}), we set $f_{\mathrm{PRF}}^{\mathrm{eq}}=400$ Hz, corresponding to $Q_{\mathrm{sym}}=250$.
\end{itemize}

Under the considered spotlight geometry, the amplitude variation is mild relative to the propagation phase, and we therefore approximate $A_m^{p,q}[k,\ell]\approx A_m$.
The number of slow-time samples $N_{\mathrm{sa}}$ is set to 512, yielding a short synthetic aperture duration of $T_{\mathrm{sa}}=1.28$ s and an aperture length $L_{\mathrm{sa}}=19.2$ m. This relatively short aperture limits the accumulation of navigation and motion errors during prolonged UAV flights. Under these configurations, the theoretical range and azimuth resolutions of the SAR system can be approximated as
\begin{equation}
		\delta_{\mathrm{rg}}  \approx \frac{c}{2B} \approx 0.375 ~\mathrm{m}, \quad
		\delta_{\mathrm{az}}  \approx \frac{\lambda R_{\mathrm{far}}}{4L_{\mathrm{sa}}} \approx 0.384~\mathrm{m}.
\end{equation}
Guided by the spatial Nyquist sampling theorem and physical diffraction limits, the continuous 100 m $\times$ 100 m target scene is discretized into a 256 $\times$ 256 grid. This specific partitioning yields a physical pixel size of 0.39 m $\times$ 0.39 m, which is deliberately configured to be commensurate with the theoretical resolutions. Such grid design ensures full utilization of ISAC system's physical resources, while avoiding the computational overhead caused by an excessively large measurement matrix resulting from overly fine grid partitioning.

\begin{figure*}[t]
	\centering
	\subfloat[Ground Truth]{%
		\includegraphics[width=0.135\textwidth]{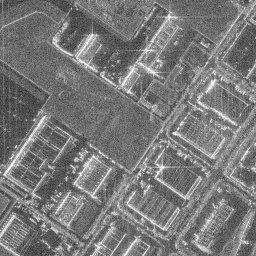}
	}
	\hfill
	\subfloat[DPS-1000]{%
		\includegraphics[width=0.135\textwidth]{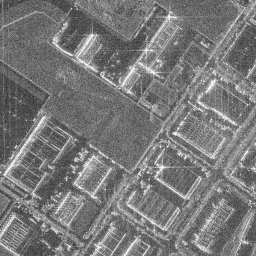}
	}
	\hfill
	\subfloat[DPS-20]{%
		\includegraphics[width=0.135\textwidth]{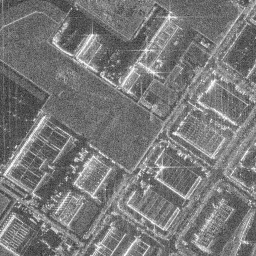}
	}
	\hfill
	\subfloat[PnP-BM3D]{%
		\includegraphics[width=0.135\textwidth]{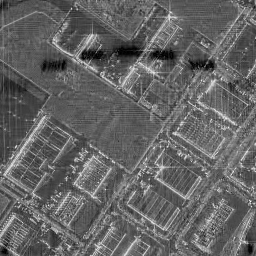}
	}
	\hfill
	\subfloat[FISTA]{%
		\includegraphics[width=0.135\textwidth]{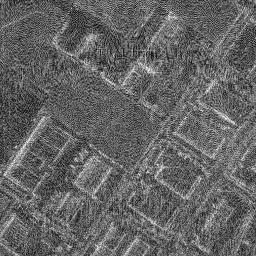}
	}
	\hfill
	\subfloat[BP]{%
		\includegraphics[width=0.135\textwidth]{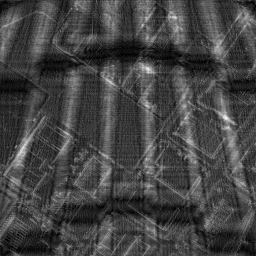}
		\label{fig:BP}
	}
	\hfill
	\subfloat[RDA]{%
		\includegraphics[width=0.135\textwidth]{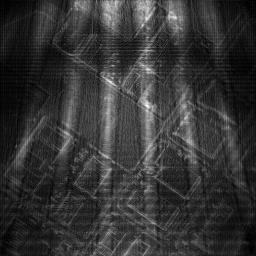}
		\label{fig:RDA}
	}
	\caption{Reconstructed SAR images of different algorithms under the ideal ULM trajectory at SNR = 20 dB.}
	\label{fig:reconstruction_comparison}
		\vspace{-3mm}
\end{figure*}

For MoCo experiments,  the actual UAV trajectory is generated by adding a 3D motion error $\Delta \mathbf{p}(\eta)$ to the reference trajectory. Following typical UAV dynamics \cite{moco1,moco2}, this error comprises low-frequency drift and high-frequency vibration $\Delta\mathbf{p}(\eta) = \Delta\mathbf{p}_{\mathrm{low}}(\eta) + \Delta\mathbf{p}_{\mathrm{high}}(\eta)$.
The low-frequency drift is modeled as $\Delta \mathbf{p}_{\mathrm{low}}(\eta) = \mathbf{c}_0+\mathbf{c}_1 \eta+\mathbf{c}_2 \eta^2$, where $\mathbf{c}_0,\mathbf{c}_1,\mathbf{c}_2 \in \mathbb{R}^3$ denote the unknown polynomial coefficients.
The high-frequency jitter is a superposition of $N_{\mathrm{vib}}$ sinusoidal vibrations, yielding
$\Delta \mathbf{p}_{\mathrm{high}}(\eta) = \sum_{l=1}^{N_{\mathrm{vib}}} \mathbf{a}_l\sin (2\pi f_l \eta + \boldsymbol{\phi}_l)$, where $\mathbf{a}_l,f_l,\boldsymbol{\phi}_l$ are the amplitude, vibration frequency, initial phase, respectively. 
The high-frequency vibration amplitude are set on the order of the signal wavelength.

We utilize the FAIR-CSAR dataset \cite{dataset} to train the SAR diffusion prior, which contains diverse typical SAR scenes such as airports and harbors. The dataset provides 12,000 training and 2,700 testing single look complex (SLC) images, originally sized at 1024 $\times$ 1024 pixels with a 1 m $\times$ 1 m physical resolution. 
The raw SLC samples are radiometrically processed and normalized to obtain complex-valued SAR reflectivity images.
They are subsequently spatially resampled and cropped to match the 256 $\times$ 256 grid required by our specific ISAC configuration. The diffusion model adopts a standard U-Net architecture adapted from DDPM \cite{DDPM}. The other relevant algorithm parameters are summarized in Table I.

The following SAR baseline methods are compared. $\textbf{RDA}$: the conventional SAR method based on the ULM assumption \cite{RDA}; $\textbf{BP}$: the geometry-based back-projection method \cite{BP}; $\textbf{FISTA}$: the compressed sensing method using hand-crafted $l_1$ sparsity priors \cite{CS_SAR2}; $\textbf{PnP}$-$\textbf{BM3D}$: an iterative optimization framework utilizing BM3D denoiser as an implicit prior \cite{pnp}. $\textbf{DPS}$-$\textbf{20}$: our proposed DDIM-accelerated DPS with $S=20$, whereas $\textbf{DPS}$-$\textbf{1000}$ denotes the full-step DDIM-accelerated DPS executing all 1000 steps, serving as the performance upper bound.
We employ peak signal-to-noise ratio ($\textbf{PSNR}$) and structural similarity index measure ($\textbf{SSIM}$) to assess the absolute intensity accuracy and spatial structure preservation, respectively.

\subsection{SAR Imaging Performance}


We first evaluate the proposed DPS-based framework under the ideal ULM trajectory.
Fig. \ref{fig:reconstruction_comparison} provides a visual comparison of the reconstructed SAR images at SNR = 20 dB.
RDA and BP suffer from pronounced sidelobes, ghosting, and structural blur, while FISTA and PnP-BM3D partially suppress these artifacts but still lose fine structural details.
This degradation is reflected by over-smoothed regions and localized artifacts, indicating the limited capability of handcrafted image priors in representing complex SAR scattering characteristics. In contrast, both DPS-20 and DPS-1000 exhibit remarkable generative capabilities, recovering substantially sharper boundaries and more faithful scattering structures, closely resembling the ground truth.

\begin{figure*}[t]
	\centering
	\begin{minipage}[t]{0.32\textwidth}
		\centering
		
		\subfloat[]{
			\includegraphics[width=\linewidth]{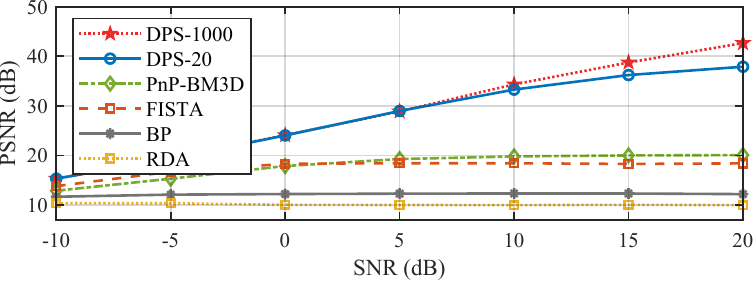}
			\label{fig:ideal_psnr}
		}
		
		\vspace{-2mm}
		
		\subfloat[]{
			\includegraphics[width=\linewidth]{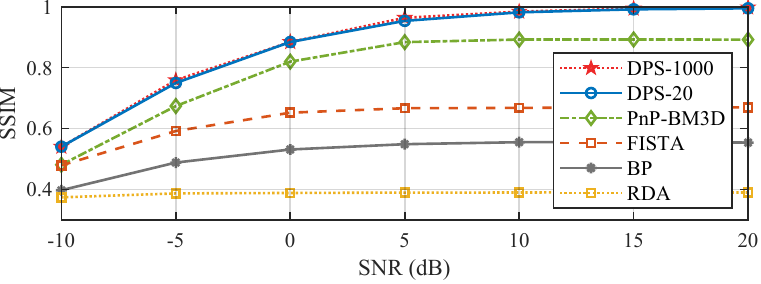}
			\label{fig:ideal_ssim}
		}
		
		\captionof{figure}{Performance of
			(a) PSNR and (b) SSIM versus SNR under the ideal ULM trajectory.
		}
		\label{fig:ideal_quantitative}
	\end{minipage}
	\hfill
	\begin{minipage}[t]{0.32\textwidth}
		\centering
		
		\subfloat[]{
			\includegraphics[width=\linewidth]{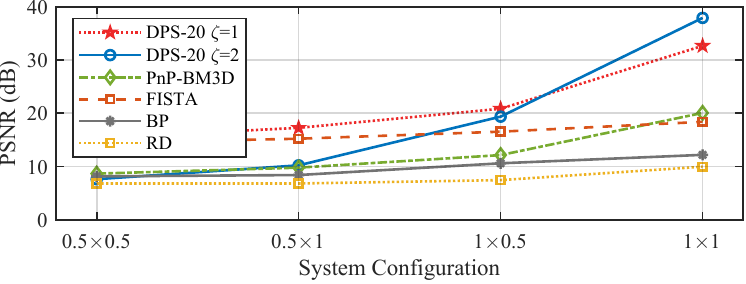}
			\label{fig:SR_PSNR}
		}
		
		\vspace{-2mm}
		
		\subfloat[]{
			\includegraphics[width=\linewidth]{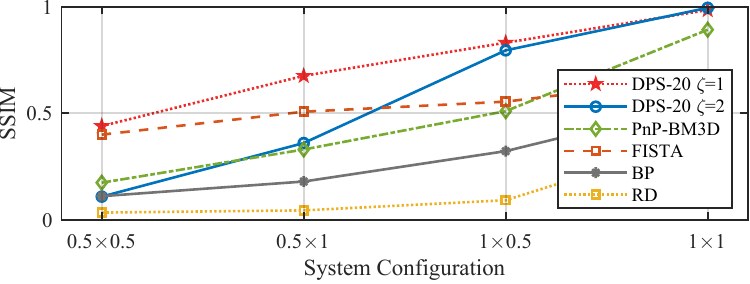}
			\label{fig:SR_SSIM}}
		\captionof{figure}{
			Comparison of (a) PSNR and (b) SSIM under \textbf{Case 1}.
		}
		\label{fig:superRe}
	\end{minipage}
	\hfill
	\begin{minipage}[t]{0.32\textwidth}
		\centering
		
		\subfloat[]{
			\includegraphics[width=\linewidth]{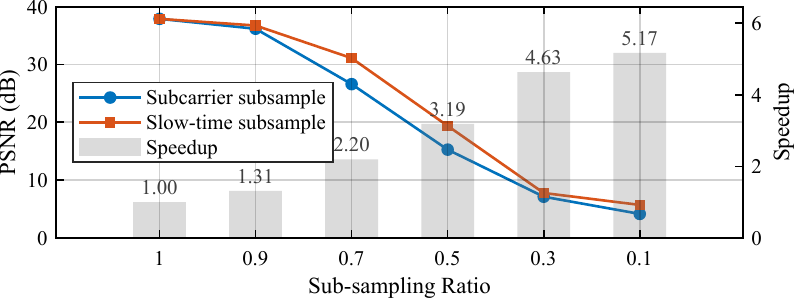}
			\label{fig:sparse_PSNR}}
		
		\vspace{-2mm}
		
		\subfloat[]{
			\includegraphics[width=\linewidth]{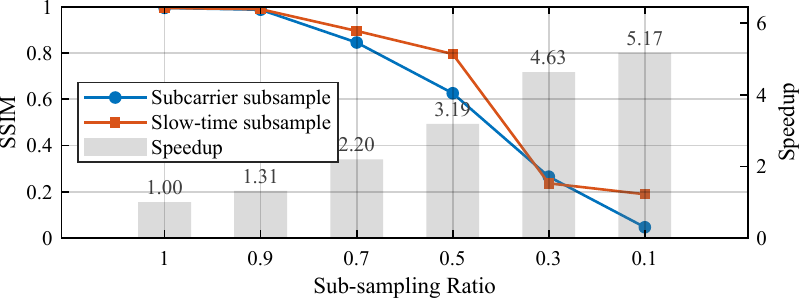}
			\label{fig:sparse_SSIM}}
		
		\captionof{figure}{
			Comparison of (a) PSNR and (b) SSIM with the corresponding speedup under \textbf{Case 2}.}
		\label{fig:sparse}
	\end{minipage}
		\vspace{-3mm}
\end{figure*}

These visual advantages are quantitatively validated in Fig. \ref{fig:ideal_quantitative}, which illustrates the PSNR and SSIM values across SNRs from -10 to 20 dB. 
RDA and BP saturate early due to their limited reconstruction capability, while FISTA and PnP-BM3D benefit from prior regularization but remain below the DPS methods. At SNR = 0 dB, both DPS variants outperform PnP-BM3D by approximately 5.7 dB in PSNR. At SNR = 20 dB, DPS-1000 and DPS-20 achieve 42.7 dB and 37.9 dB, respectively, with SSIM values approaching 1. Notably, DPS-20 closely matches DPS-1000 in the low-to-moderate SNR regime and maintains nearly identical SSIM across the entire SNR range, while requiring only 20 DDIM sampling steps. These results demonstrate that DDIM acceleration preserves most of the reconstruction quality of full-step DPS while substantially reducing inference cost.



\subsection{Super-Resolution Performance}
In practical low-altitude ISAC systems, communication constraints and hardware limitations prevent sensing from acquiring ideal full-resolution or fully sampled observations. To evaluate the capability of diffusion prior to break physical limitations, we construct two experiments at SNR = 20 dB to investigate the effects of physical resolution coarsening and sub-Nyquist sparse sampling.

$\textbf{Case 1: Physical Resolution Coarsening.}$ We first reduce the available bandwidth and synthetic aperture duration to degrade the physical resolution.
In Fig. \ref{fig:superRe}, the 1$\times$1 represents the origin settings where the physical resolution perfectly matches the pixel grid 0.39 m. The 0.5$\times$1 halves the bandwidth, degrading the theoretical range resolution to $\sim$0.75 m, considerably coarser than the pixel size. The configuration 1$\times$0.5 halves the synthetic aperture, while 0.5$\times$0.5 halves both. 
Such severe mismatch between the theoretical system resolution and the pixel size leads to an ill-posed inverse problem, classical matched filtering BP completely fails. While FISTA maintains certain super-resolution capabilities due to its sparsity prior, its comprehensive performance remains limited. 
Instead, DPS-20 remains effective under aperture truncation, achieving 21.3 dB PSNR and 0.81 SSIM, nearly 4.0 dB higher than FISTA. 
However, DPS-20 is sensitive to bandwidth truncation, initially yielding sub-optimal performance under 0.5$\times$1. 
To address this, we reduce the hyperparameter $\zeta$, which controls the adaptive step size $\rho_{i_{s}}$ of the likelihood score in (\ref{III-C-eq:rho_step_size}), from its default value of 2 to 1. This operation dynamically elevates the relative weight of diffusion prior during the reverse sampling process, compensating for the severe information loss in $\widetilde{\boldsymbol{\Phi}}^{\mathrm{ref}}$. DPS-20 with $\zeta=1$ significantly outperforms all baselines under severe truncation conditions, yielding a PSNR improvement of nearly 10 dB compared to $\zeta=2$. These results demonstrate that the learned diffusion prior effectively regularizes the ill-posed inverse problem and enables robust super-resolution reconstruction under degraded physical resolution.

$\textbf{Case 2: Sub-Nyquist Sparse Sampling.}$ Keeping the theoretical resolution (preserving the total bandwidth and aperture length), we apply sub-sampling to the subcarriers and slow-time measurements to simulate data loss cased by practical ISAC challenges such as communication scheduling or target occlusions. While traditional algorithms suffer from severe sidelobes and grating lobes under sub-Nyquist conditions, the DPS-based framework exhibits remarkable robustness. 
In Fig. \ref{fig:sparse}, we evaluate the performance of DPS-20 under these conditions. Remarkably, when either subcarriers or slow-time are subsampled at a ratio of 0.9, the imaging quality remains unaffected. Even as the subsampling ratio drastically drops to 0.7, the PSNR exhibits an acceptable degradation of within 10 dB, while the SSIM firmly stabilizes above 0.8. This resilience further demonstrates that the diffusion prior effectively interpolates the missing data manifold, thereby endowing DPS with inherent sparse super-resolution capabilities. Moreover, sparse sampling reduces the dimension of the data-consistency gradient $\mathbf{g}_{i_{s}}$ computation. As indicated by the speedup in Fig. \ref{fig:sparse}, discarding 30$\%$ of the measurements directly yields an approximately 2.2$\times$ acceleration in generative inference, demonstrating that moderate undersampling can substantially improve inference efficiency with limited reconstruction degradation.

\begin{figure}[t]
	\centering
	\includegraphics[width=0.8\linewidth]{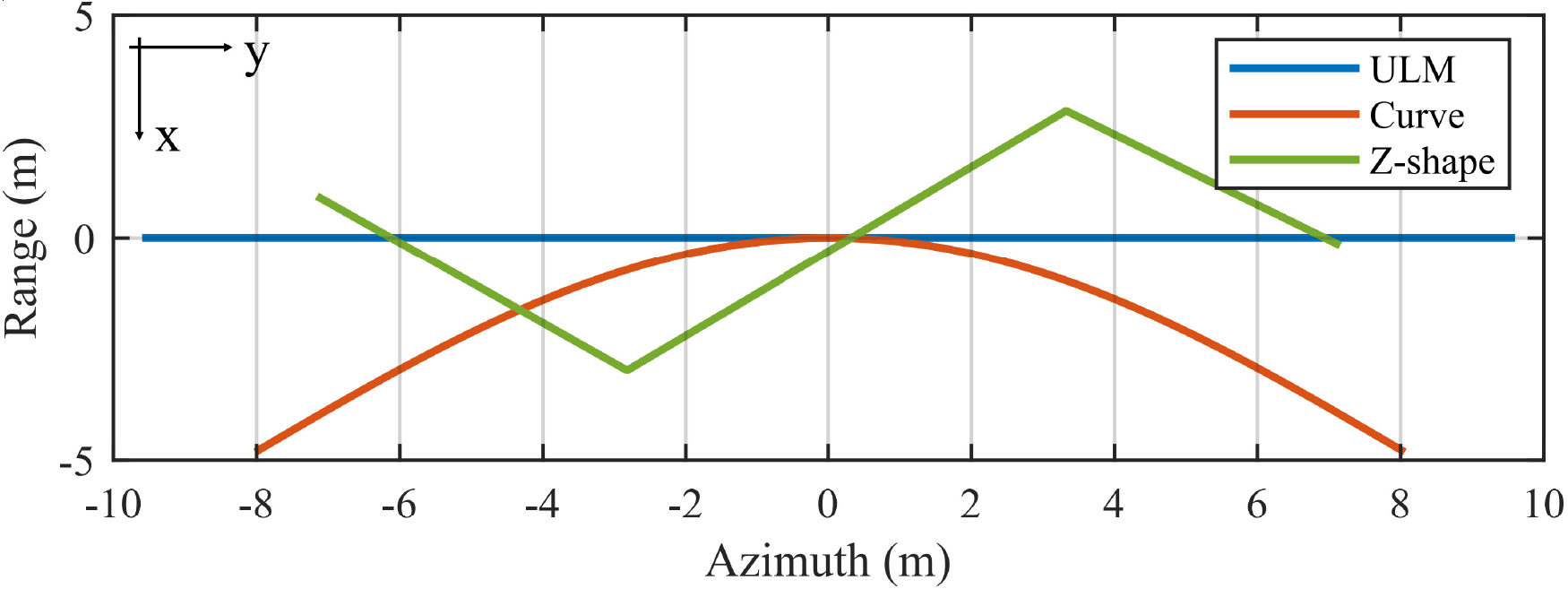}
	\vspace{-2mm}
	\caption{Reference UAV trajectories used for robustness evaluation.}
	\label{fig:three-trajs}
	\vspace{-3mm}
\end{figure}
\vspace{-2mm}

\begin{figure}[t]
	\centering
	\subfloat[]{
		\includegraphics[width=0.4\textwidth]{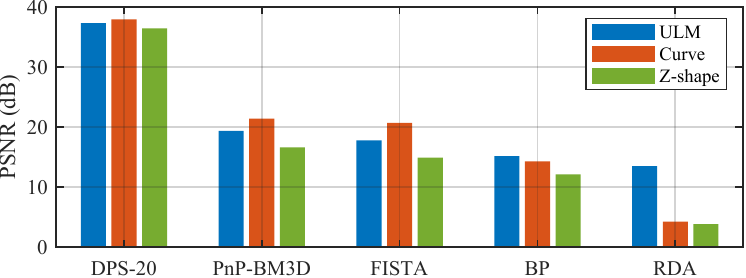}
		\label{fig:tablepsnr}}

	\subfloat[]{
		\includegraphics[width=0.4\textwidth]{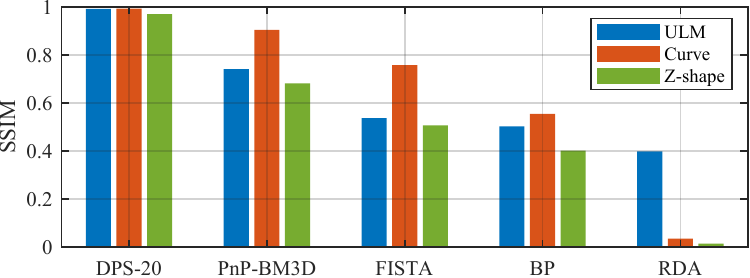}
		\label{fig:tablessim}}
	\caption{Performance of (a) PSNR and (b) SSIM across diverse reference trajectories.}
	\label{fig:muTJ}
	\vspace{-3mm}
\end{figure}

\subsection{Robustness to Diverse Trajectories}
We further evaluate the robustness of the proposed framework to different known UAV reference trajectories at SNR = 20 dB. As shown in Fig. \ref{fig:three-trajs}, three representative trajectories are considered: an ideal ULM trajectory, a smooth curved trajectory, and a Z-shape trajectory. All trajectories have the same path length and identical $f_{\mathrm{PRF}}^{\mathrm{eq}}$, resulting in the same synthetic aperture duration and number of uniform slow-time samples.

As illustrated in Fig. \ref{fig:muTJ}, RDA performs reasonably under ULM but degrades sharply under the curved and Z-shape trajectories, consistent with the ULM-model mismatch analyzed in Section II-C. In contrast, BP exploits the exact propagation geometry, while FISTA, PnP-BM3D, and DPS-20 reconstruct with the corresponding trajectory-dependent operator $\widetilde{\boldsymbol{\Phi}}^{\mathrm{ref}}$, thereby avoiding severe model mismatch-induced defocusing.
Nevertheless, their performance still varies with trajectory because different viewing-angle evolutions yield different effective synthetic apertures and different inverse problem conditioning. Therefore, some geometry-aware methods improve under the curved trajectory but degrade under the Z-shape trajectory.
Despite these geometry-induced variations, DPS-20 remains highly stable across all three trajectories, with only 1.50 dB PSNR variation and less than 0.024 SSIM variation, substantially smaller than those of the conventional baselines. 
Even under the Z-shape trajectory, DPS-20 retains 36.39 dB PSNR and 0.969 SSIM, whereas the best baseline, PnP-BM3D, achieves only 16.62 dB and 0.681.
More importantly, this robustness also reflects the empirical cross-trajectory generalization of the proposed framework. Since the diffusion prior is learned solely from SAR reflectivity data and is decoupled from the sensing geometry, changing UAV trajectory only requires updating $\widetilde{\boldsymbol{\Phi}}^{\mathrm{ref}}$, while the same pre-trained diffusion model can be reused without trajectory-specific retraining.

\vspace{-3mm}

\subsection{Motion Compensation Performance}
We evaluate the comprehensive performance of the proposed joint DPS and MoCo framework when unknown three-dimensional position errors are superimposed on the ideal ULM reference trajectory. The root-mean-square magnitude of the injected position error $\Delta\mathbf{p}(\eta)$ is set to 0.5$\lambda$.
Table \ref{tab:moco_results} reports the reconstruction PSNR and SSIM at SNR = 20 dB, together with the position-error root-mean-square error (RMSE), mean 3D Euclidean error (MEE), and the percentage of slow-time samples with residual errors below $\lambda/16$ as position-error estimation metrics. The MEE measures the average spatial deviation in 3D space, whereas the $< \lambda/16$ ratio characterizes the accuracy required for coherent focusing. Notably, since DPS-1000 executes full 1000 diffusion sampling steps,  significantly higher than baselines, it is included only as a performance upper bound.

In Table \ref{tab:moco_results}, FISTA and PnP-BM3D also alternately update the image and position errors. However, their relatively weak priors lead to inaccurate image estimates in early iterations, which can misguide subsequent position-error refinement and cause error accumulation. For example, PnP-BM3D yields an RMSE and MEE of 2.94 cm and 3.19 cm, respectively, both comparable to 0.5$\lambda$, while the percentage below $\lambda/16$ is nearly zero, indicating substantial deviation from the true motion. The single-step BP and RDA methods avoid such iterative error accumulation and achieve lower RMSEs of 0.24 cm and 0.63 cm, but their reconstruction capability remains limited, with PSNRs of only 16.29 dB and 14.59 dB. In contrast, DPS-20 achieves 23.65 dB PSNR and 0.91 SSIM while reducing the position-error RMSE and MEE to 0.19 cm and 0.27 cm, respectively, thereby achieving both high-quality reconstruction and accurate motion compensation.

\begin{table}[t]
	\centering
	\caption{Joint DPS and MoCo performance at SNR = 20 $\mathrm{dB}$.}
	\label{tab:moco_results}
	\begin{tabular}{cccccc}
		\toprule
		Method & \makecell{PSNR\\(dB)} $\uparrow$ & SSIM $\uparrow$ 
		& \makecell{RMSE\\ (cm)} $\downarrow$ 
		& \makecell{MEE\\ (cm)} $\downarrow$ 
		& \makecell{$<\lambda/16$\\ (\%)} $\uparrow$ \\
		\midrule
		RDA      & 14.59 & 0.43 & 0.63 & 0.61 & 66.41 \\
		BP       & 16.29 & 0.55 & 0.24 & 0.37 & 58.59 \\
		FISTA    & 19.63 & 0.68 & 0.58 & 0.55 & 66.79 \\
		PnP-BM3D & 19.14 & 0.61 & 2.94 & 3.19 & 7.18e-03 \\
		DPS-20      &\textbf{23.65} & \textbf{0.91} & \textbf{0.19} & \textbf{0.27} & \textbf{80.26} \\
		\midrule
		DPS-1000  & 29.24 & 0.99 & 0.15 & 0.23 & 86.62 \\
		\bottomrule
	\end{tabular}
		\vspace{-3mm}
\end{table}
 
\begin{figure}[t]
	\centering
	\includegraphics[width=0.95\linewidth]{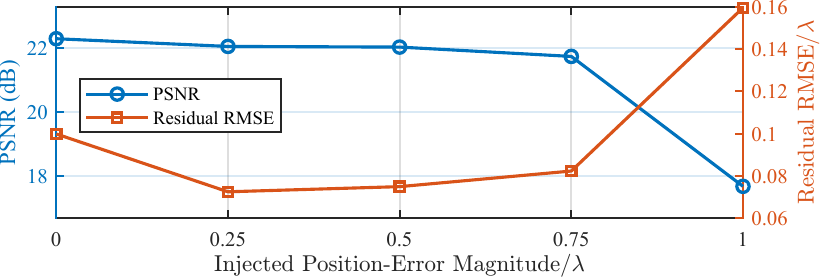}
	\vspace{-2mm}
	\caption{Performance of PSNR and residual position-error RMSE versus the input position-error magnitude.}
	\label{fig:diff_position_error}
		\vspace{-3mm}
\end{figure}

We further evaluate the robustness of the joint DPS and MoCo framework to different position-perturbation levels by proportionally scaling the injected position-error magnitude. As shown in Fig. \ref{fig:diff_position_error}, when the injected position error increases to 0.75$\lambda$, the residual RMSE remains low and the PSNR decreases only gradually, indicating effective mitigation of the measurement operator mismatch over this range. When the error further increases to 1$\lambda$, the performance degrades, but the residual RMSE remains around 0.16$\lambda$, demonstrating that meaningful compensation is still retained under stronger flight perturbations. Notably, the small residual at zero injected error arises because MoCo remains active, where noise and intermediate estimation errors can still induce nonzero updates, resulting in an estimation floor.

\begin{table}[t]
	\caption{Ablation study of DCT reparameterization and coarse-to-fine scheduling.}
	\label{tab:dct_c2f_ablation}
	\centering
	\footnotesize
	\setlength{\tabcolsep}{3.5pt}
	\begin{tabular}{ccccccc}
		\toprule
		Method
		& \makecell{RMSE$_{25}$\\ (cm)}
		& \makecell{RMSE$_{50}$\\ (cm)}
		& \makecell{RMSE$_{75}$ \\(cm)}
		& \makecell{Final RMSE \\(cm)}
		& \makecell{PSNR \\ (dB)}
		
		\\
		\midrule
		Time-domain 
		& 2.89 
		& 14.86 
		& 21.12 
		& 34.14 
		& 10.57 
		
		\\
		
		Fixed-small DCT 
		& 0.64 
		& 0.58 
		& 0.48 
		& 0.27 
		& 16.84 
		
		\\
		
		Fixed-large DCT 
		& 3.30 
		& 6.09 
		& 7.59 
		& 12.01 
		& 10.65 
		
		\\
		
		Proposed C2F-DCT
		& \textbf{0.50}
		& \textbf{0.38}
		& \textbf{0.32}
		& \textbf{0.19}
		& \textbf{23.65}
		
		\\
		\bottomrule
	\end{tabular}
	\vspace{-3mm}
\end{table}

\vspace{-3mm}

\subsection{Ablation Study}
We conduct an ablation study on the DCT reparameterization and C2F spectral scheduling  in Section IV-B. Four position-error optimization strategies are compared, including direct time-domain optimization, fixed-small DCT, fixed-large DCT, and the proposed C2F-DCT. In Table \ref{tab:dct_c2f_ablation}, $\mathrm{RMSE}_{*}$ denotes the cumulative average position-error RMSE over the first $*\%$ reverse diffusion steps, where $*\in\{25,50,75\}$, characterizing the stability of the joint DPS and MoCo process. The Final RMSE measures the final position-error estimation accuracy.

As shown in Table \ref{tab:dct_c2f_ablation}, direct time-domain optimization becomes progressively unstable, with the RMSE increasing from 2.89 cm to a final value of 34.14 cm. Fixed-large DCT also suffers from early exposure to excessive high-frequency degrees of freedom, causing the RMSE to increase from 3.30 cm to 12.01 cm. In contrast, fixed-small DCT stabilizes the refinement through stronger low-frequency regularization and achieves a final RMSE of 0.27 cm, but its PSNR is limited to 16.84 dB, indicating insufficient representation capability for fine-scale position errors at later stages. The proposed C2F-DCT maintains a steadily decreasing RMSE and reaches a final value of 0.19 cm, together with the highest PSNR of 23.65 dB. These results validate the intended C2F mechanism, where a compact DCT subspace suppresses unstable updates in the early stages and is progressively expanded as the reflectivity estimate becomes more reliable, thereby balancing optimization stability and fine-scale error representation.

\vspace{-2mm}
\section{Conclusion}
This paper developed a diffusion-based framework for super-resolution ISAC SAR imaging under non-ideal UAV motion. For known arbitrary reference trajectories, an exact trajectory-dependent measurement operator was combined with a learned diffusion prior through DPS, avoiding ULM-based decoupling approximations while providing strong regularization for ill-posed fine-grid reconstruction. DDIM acceleration substantially reduced the inference cost with limited performance loss. For unknown position errors, the proposed joint DPS and MoCo framework iteratively refined the SAR reflectivity and position errors, while DCT reparameterization and C2F spectral scheduling improved the stability of position-error estimation. Numerical results demonstrated robust reconstruction across diverse trajectories, super-resolution under physical-resolution truncation, and accurate motion compensation under severe position perturbations. These results establish a unified framework for high-fidelity SAR imaging under practical non-ideal motion in low-altitude ISAC systems.

\bibliographystyle{IEEEtran} 
\bibliography{DPS_SAR}

\end{document}